\PassOptionsToPackage{table}{xcolor}
\documentclass[a4paper, fleqn]{cas-sc}

\usepackage{natbib}
\usepackage{listings}
\usepackage{subfig}
\usepackage{titlesec}
\usepackage{lipsum}
\usepackage{setspace}
\usepackage{stackengine}
\usepackage{float}
\usepackage{array}
\usepackage{graphicx}
\usepackage{booktabs}
\usepackage{amsmath}
\usepackage{placeins}
\usepackage{makecell}
\usepackage{tabularx}
\usepackage[tableposition=top]{caption}

\renewcommand{\arraystretch}{1.5}
\restylefloat{table}
\definecolor{dkgreen}{rgb}{0,0.6,0}
\definecolor{mauve}{rgb}{0.58,0,0.82}
\definecolor{darkgraycustom}{HTML}{4D4D4D}
\definecolor{brightred}{HTML}{FF0000}

\titleformat{\section}{\normalfont\Large\bfseries}{\thesection}{1em}{}
\titleformat{\subsection}{\normalfont\large\bfseries}{\thesubsection}{1em}{}
\titleformat{\subsubsection}{\normalfont\normalsize\bfseries}{\thesubsubsection}{1em}{}
\titlespacing*{\section}{0pt}{5mm}{3mm}
\titlespacing*{\subsection}{0pt}{5mm}{3mm}

\begin{document}
\let\WriteBookmarks\relax
\def\floatpagepagefraction{1}
\def\textpagefraction{.001}
\shorttitle{Algorithm-driven Development}
\shortauthors{Philippe Jawish et~al.}

\title [mode = title]{Algorithm-driven Development: A Proactive Approach to Improving Software Quality and Reducing Defects}

\author[]{\texorpdfstring{\hyperlink{philippe_bio}{Philippe Jawish}}{Philippe Jawish}}
\author[]{\texorpdfstring{\hyperlink{pierre_bio}{Pierre Evrard}}{Pierre Evrard}}
\author[]{\texorpdfstring{\hyperlink{alexandre_bio}{Alexandre Lemerle}}{Alexandre Lemerle}}
\author[]{\texorpdfstring{\hyperlink{adrian_bio}{Adrian Genin}}{Adrian Genin}}
\author[]{\texorpdfstring{\hyperlink{layal_bio}{Layal Dergham}}{Layal Dergham}}
\author[]{\texorpdfstring{\hyperlink{severin_bio}{Séverin Lanfranchi}}{Séverin Lanfranchi}}

\affiliation[]{organization={Dassault Systèmes},
                addressline={10 Rue Marcel Dassault}, 
                city={Vélizy-Villacoublay},
                postcode={78140},
                country={France}}
         
\begin{abstract}
Ensuring software quality while meeting deadlines and adapting to evolving requirements is a persistent challenge in software engineering practice. This paper introduces Algorithm-Driven Development (ADD), a methodology developed from industrial practice to address recurring challenges in translating requirements into reliable, testable, and maintainable software behavior. ADD translates requirements into algorithmic flowcharts from which acceptance tests are systematically derived. These flowcharts serve both as specification artifacts and as technical blueprints, supporting shared understanding between stakeholders and developers. By linking requirement modeling with automated test generation, ADD provides systematic coverage of functional scenarios, including edge cases, from the outset of development. The approach was evaluated over a four-year period within an industrial project at Dassault Systèmes, involving two development teams, 22,444 lines of production code for Team~1, and 157 APIs analyzed for Team~2. The evaluation combined longitudinal quality and delivery indicators with a comparative analysis of ADD, TDD, and test-last development practices across API functions of different complexity levels. Empirical data collected from internal lifecycle management and CI/CD systems show that ADD supported sustained code coverage above 95\%, low defect density in both QA and post-release phases, and a stable delivery cadence. These findings provide evidence of ADD’s potential to strengthen the connection between requirements, testing, and implementation in industrial software development contexts.

\end{abstract}

\begin{keywords}
Test-driven Development\sep 
Behavior-driven Development\sep 
Acceptance test-driven development\sep 
Algorithm-driven Development\sep 
ADD
\end{keywords}

\maketitle

\setlength{\parskip}{6pt}

\section{Introduction}

In the field of modern computer science, a variety of methods and principles have been developed to optimize the software development process while enhancing the quality of the final product \citep{al2020agile}. However, these approaches often prioritize quality at the expense of other critical factors, such as time constraints, business imperatives, and customer expectations. Among these, Test-Driven Development (TDD) and Behavior-Driven Development (BDD) are recognized for their effectiveness in improving software quality \citep{abushama2021effect}. TDD, in particular, allows developers to write tests before coding, ensuring functionality and reducing defects.

Yet, while TDD and BDD have brought improvements, they also reveal limitations, particularly in projects involving complex systems or multiple dependencies. Specifically, TDD and BDD lack a structured mechanism for proactively addressing system complexity, which becomes a significant problem as software systems grow in scale and interdependence \citep{ganja2023investigation}. Additionally, BDD involves extensive collaboration between stakeholders, such as business analysts, Quality Assurance team (QA), and developers, which can be time-consuming and difficult to maintain consistently \citep{farooq2023behavior}, especially in fast-paced projects. The failure to anticipate all possible scenarios and dependencies results in development inefficiencies, impacting timelines, quality, customer satisfaction, and ultimately leading to financial losses for the company \citep{Krasner2020}. 

These issues are not limited to abstract theoretical risks; they carry a measurable economic burden. In 2020, the total cost of poor software quality in the United States alone was estimated at \$2.08 trillion, of which \$1.56 trillion was attributed to operational defects such as bugs, system failures, and user-facing issues \citep{Krasner2020}. Beyond this economic impact, empirical evidence also shows that software defects remain prevalent across development contexts. Reported defect-density values vary according to factors such as programming language, project context, defect definition, and measurement phase. For example, \citet{shah2012overview} report an average defect density of 7.47 defects/KLOC across 109 projects and 5.9 defects/KLOC for Java projects. Together, these economic and empirical observations underline the need for improved development practices that address software quality in a structured and proactive manner.

The inefficiency and difficulty of managing complexity in large-scale software projects have highlighted the need for a more structured and proactive approach \citep{kasauli2021requirements}. To address these challenges, this paper introduces a new methodology called Algorithm-driven Development (ADD). ADD introduces a proactive approach that enhances TDD by focusing on creating algorithmic flowcharts before the implementation of tests and code. By adopting a mathematical and logic-driven approach, ADD transforms client requirements into detailed algorithms, providing a clear, visual representation of possible scenarios, including edge cases, which are defined as infrequent or extreme input or usage conditions that lie outside typical execution paths and are often overlooked in traditional specifications. This structured preparation aims to improve test design by promoting logical coverage and completeness, while contributing to software robustness by grounding tests in a detailed understanding of system behavior.

However, ADD does not aim to model entire system interactions or high-level business processes. Instead, it operates at the technical specification level, focusing on defining the behavior of specific components, such as an Application Programming Interface (API), a service, or a User Interface (UI) element. Each algorithm represents the internal logic of a distinct feature rather than the full client requirements, ensuring a detailed and precise breakdown of functionality at a granular level.
ADD contributes to the Shift Left approach \citep{vaddadi2023shift} and supports continuous quality by enabling the early identification and resolution of potential issues before coding begins. It helps reduce late-stage debugging, improve test coverage, and promote predictability and maintainability in software development.

This work is guided by three central research questions:
\begin{enumerate}
\item How can client needs be addressed comprehensively while anticipating relevant edge cases?
\item What relationship can be observed between the development approach, software defect rates, and overall product reliability?
\item How can delivery timelines be better managed to align with project constraints?
\end{enumerate}  

The recurring challenges raised by these questions motivated the development of ADD, designed to respond to the difficulties encountered in agile business contexts. ADD seeks to support software development by balancing quality considerations with constraints such as time, business requirements, and the need for flexibility in agile environments \citep{thesing2021agile}. By emphasizing algorithmic design early in the development process, ADD facilitates the anticipation of complexity and fosters stakeholder alignment throughout the project, contributing to improved consistency and quality in outcomes.

The remainder of this article is organized as follows. Section~\ref{sec:relatework} reviews related development and testing approaches, including TDD, BDD, and model-based testing, and discusses the practical gaps that motivated the development of ADD. Section~\ref{sec:ADD} presents the ADD methodology, including its core principles, process, supporting algorithms for test selection and derivation, and integration into Agile and DevOps workflows. Section~\ref{sec:resultsanddiscussion} reports empirical observations from two industrial teams: the first applied ADD over several years within a complex system, while the second integrated ADD into an existing critical service initially developed using test-last and TDD practices. The section analyzes quality, defect, and delivery indicators, and discusses contextual factors and limitations. Finally, Section~\ref{sec:conclusion} summarizes the main findings, consolidates lessons learned for practitioners, and outlines directions for future research and broader applications of ADD.

\section{Related Work and Motivation for ADD}
\label{sec:relatework}

Software engineering has introduced numerous methodologies to improve software quality, ensure requirement correctness, and reduce defect rates. Among the most widely adopted approaches are TDD, Acceptance Test-Driven Development (ATDD), and BDD, which emphasize early validation through testing and close alignment between requirements and implementation.

In parallel, model-based and graph-based testing techniques provide systematic mechanisms for deriving test cases from structured representations of system behavior, such as control-flow graphs and behavioral models. These approaches aim to improve coverage, ensure consistency between requirements and implementation, and reduce defects through systematic test generation.

This section reviews these existing methodologies and testing approaches to identify the practical gaps that motivated ADD. Since ADD is introduced in this paper, references to ADD in this section are used only to position the rationale of the proposed method; the methodology itself is described in detail in Section~\ref{sec:ADD}.

\subsection{Test-driven Development}

\begin{figure}[h]
	\centering
	\includegraphics[width=.3\textwidth]{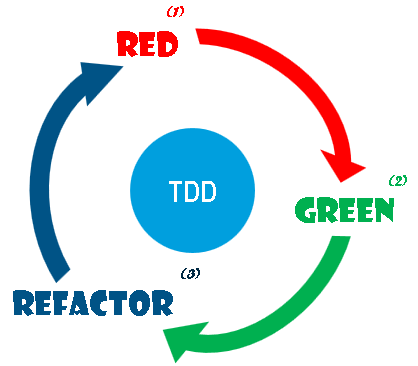}
     \caption{Test-driven Development life cycle}
     \label{fig:tdd}
\end{figure}

TDD is a software development methodology in which tests are written before implementing the corresponding functionality. Originally introduced in the context of iterative and incremental development \citep{beck2022test}, TDD follows an iterative development cycle based on the "baby steps" principle \citep{karac2018we}, consisting of three phases: RED, GREEN, and REFACTOR. Figure~\ref{fig:tdd} illustrates the TDD development cycle.

In the RED phase, a test is written to describe the expected behavior of a specific functionality. This test must compile but initially fail, as the corresponding production code has not yet been implemented. Only minimal structural elements, such as method signatures, may be introduced at this stage. The objective of this phase is to formally capture the expected behavior before implementation begins.

In the GREEN phase, the minimal production code required to pass the test is implemented. This step focuses exclusively on satisfying the test requirements without introducing additional functionality. Once the test passes successfully, the development proceeds to the REFACTOR phase.

During the REFACTOR phase, the internal structure of the code is improved to enhance readability, maintainability, and performance, while preserving its external behavior. Existing tests serve as regression safeguards to ensure that functional correctness is maintained throughout code modifications \citep{beck2022test}.

Empirical studies have demonstrated that TDD improves software reliability and facilitates regression detection by enforcing systematic validation throughout the development process \citep{bissi2016effects}. Systematic reviews have also shown that TDD improves test coverage and defect detection effectiveness at the unit level, contributing to improved software quality \citep{kollanus2011critical}. Meta-analyses of empirical studies further indicate that TDD improves external software quality, although its impact on development productivity may vary depending on project context and developer experience \citep{rafique2012effects}. In addition, improvements in test quality and defect detection effectiveness have been consistently observed when TDD is applied rigorously, particularly for component-level validation \citep{munir2014considering}.

Despite these benefits, several limitations of TDD have been identified in the literature, particularly when applied to complex systems. TDD primarily focuses on validating individual units of functionality and does not inherently provide explicit mechanisms for modeling overall system behavior or architectural structure \citep{kollanus2011critical}. As system complexity increases, identifying appropriate test scenarios and ensuring comprehensive behavioral coverage becomes increasingly difficult, requiring significant manual effort and expertise. In large-scale systems, architectural decisions often emerge incrementally during implementation, which may introduce structural constraints and increase refactoring effort \citep{staegemann2023challenges}. Furthermore, the emphasis on verifying implementation correctness rather than explicitly modeling expected system behavior may result in gaps between tested functionality and expected client behavior \citep{parsa2025testability}.

These observations indicate that while TDD is highly effective for ensuring correctness and reliability at the component level, it does not explicitly provide mechanisms for modeling execution logic or guiding architectural design in complex systems. This limitation motivates the exploration of approaches that introduce explicit behavioral modeling and systematic representation of execution logic earlier in the development process.

These limitations motivated the design of ADD around explicit behavioral modeling before implementation. Rather than relying only on tests written during implementation, ADD was developed to make execution paths and system behavior explicit earlier in the development process. The resulting methodology is presented in detail in Section~\ref{sec:ADD}.

To position ADD with respect to existing practices, it is useful to examine BDD, which extends test-driven principles beyond unit-level validation by emphasizing behavioral specification, acceptance criteria, and stakeholder alignment. Before introducing BDD in detail, the following section first discusses ATDD, since ATDD provides the acceptance-test foundation on which BDD builds.

\subsection{Acceptance Test-driven Development, double loop}

ATDD is an advanced methodology of TDD that focuses on tests that translate requested
specifications \citep{hoffmann2014applying}. These tests are considered higher-level since they describe a complete scenario
of the specification. In an agile context, acceptance tests play an essential role in ensuring that the final product meets stakeholder needs, fostering a shared understanding of requirements, and facilitating continuous validation of features throughout each iteration \citep{bjarnason2016multi}. They are mostly regarded as integration tests. This methodology is inspired
by the TDD London School approach, which involves decomposing a complex functionality to
determine all its functional dependencies \citep{Cyrille2022}.

\subsubsection{TDD London School}

\begin{figure}[h]
\footnotesize
\centering
\stackunder[1pt]{\includegraphics[width=6.5cm]{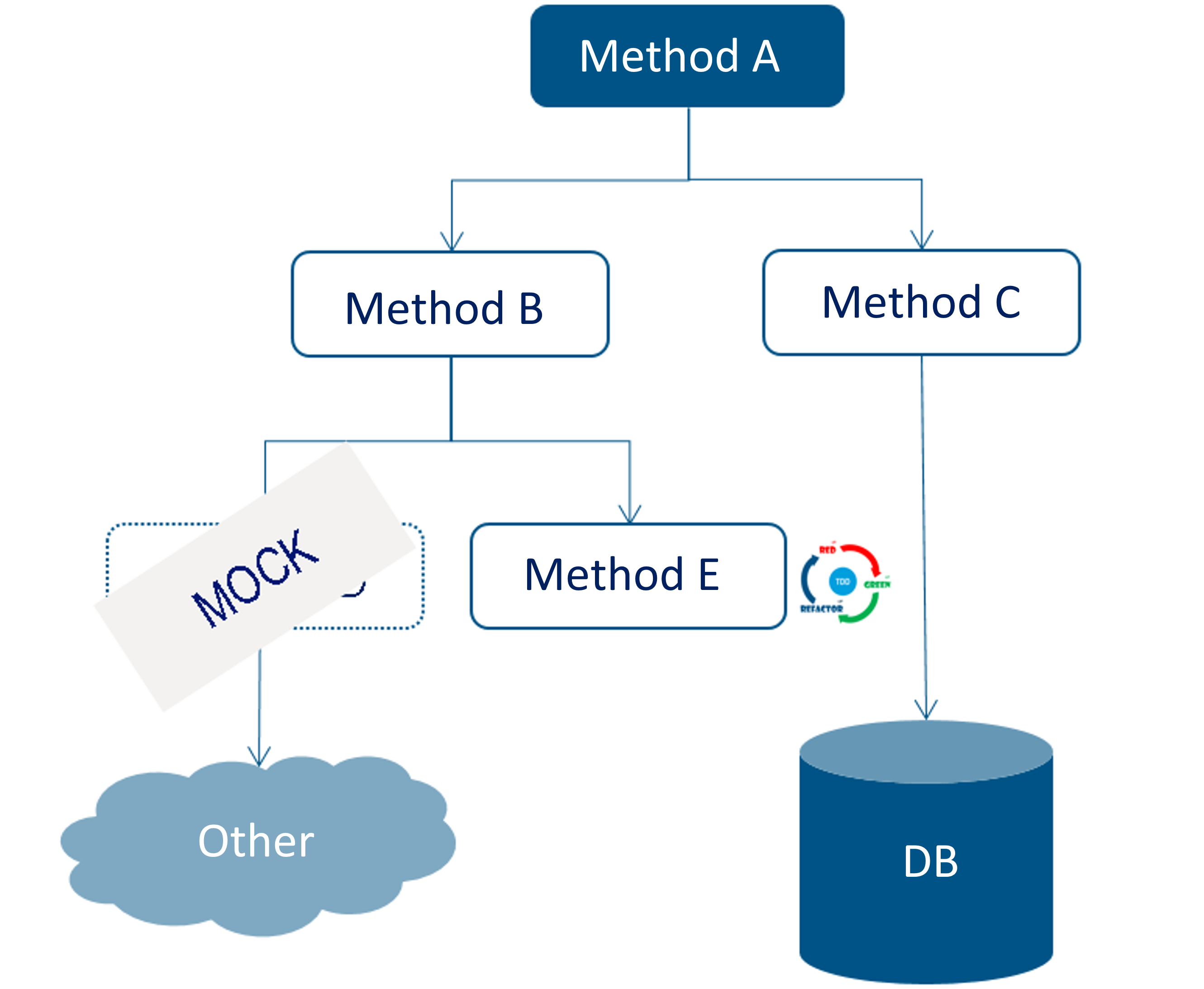}}{Step 5}
\hspace{1cm}
\stackunder[1pt]{\includegraphics[width=6.5cm]{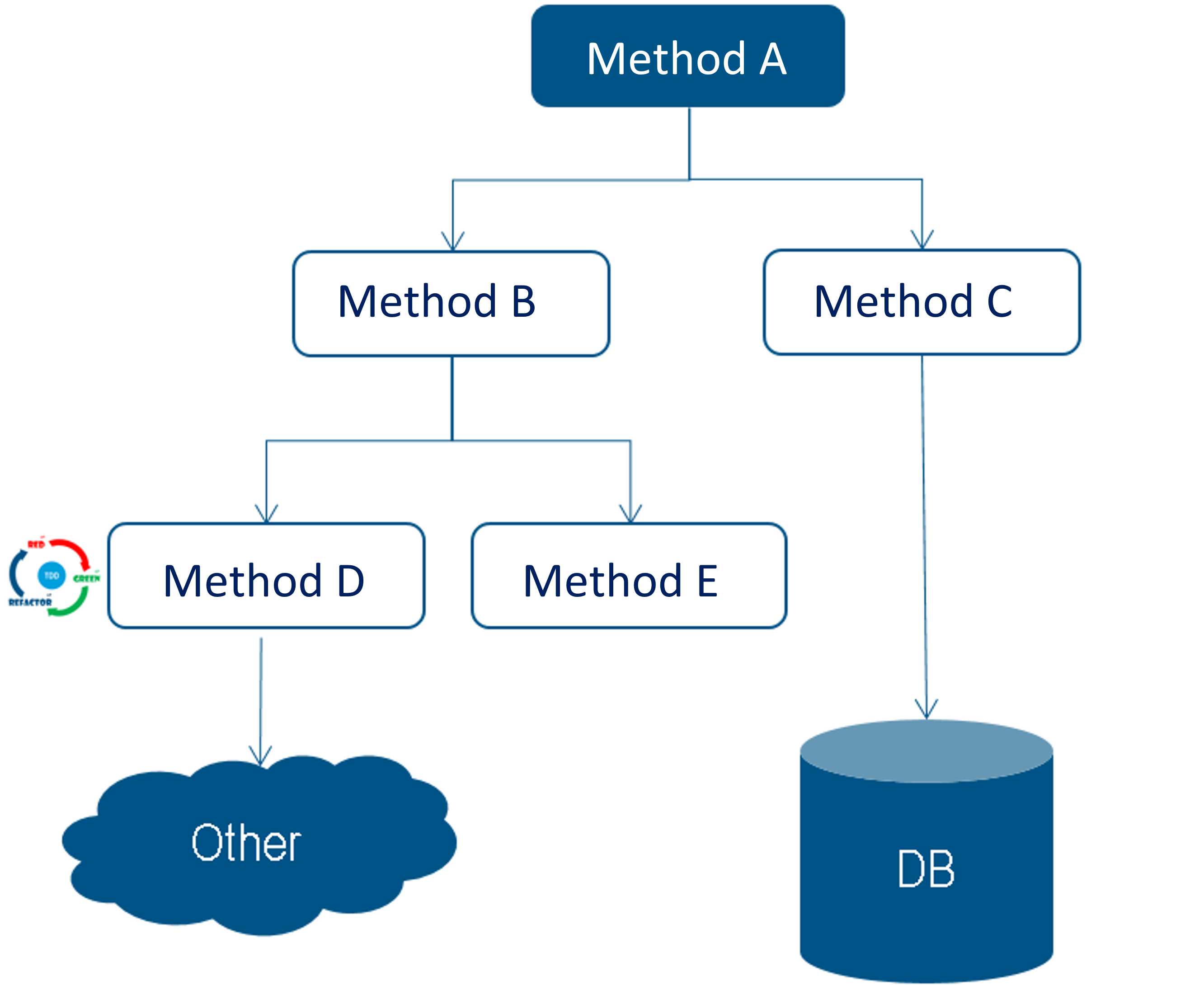}}{Step 6}
     \caption{Remaining steps of the Test-driven Development London School methodology}
     \label{fig:stepsatdd}
\end{figure}

This functional decomposition allows for better structuring of the code following the
architecture of the application. Here is an example of functional decomposition where Method
A is the main method. Once the decomposition is done, you simply develop the main method using TDD while
replacing all functional dependencies (sub-methods) with substitutes \citep{Cyrille2022}. The steps 1 and 2 of Figure~\ref{fig:firststepsatdd} illustrates how to start the development:
\begin{figure}[h]
\footnotesize
\centering
\stackunder[5pt]{\includegraphics[width=6.5cm]{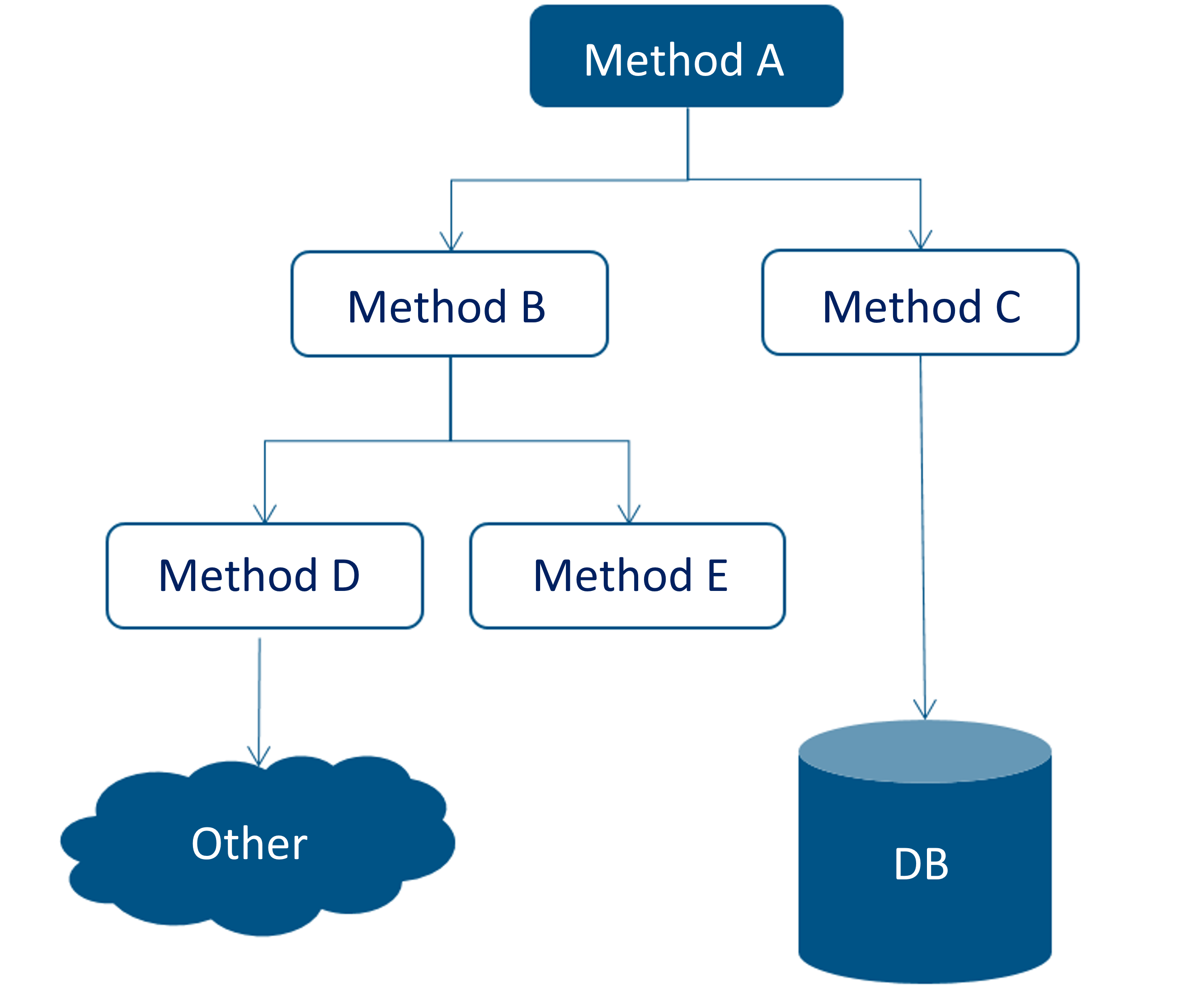}}{Step 1}
\hspace{1cm}
\stackunder[5pt]{\includegraphics[width=6.5cm]{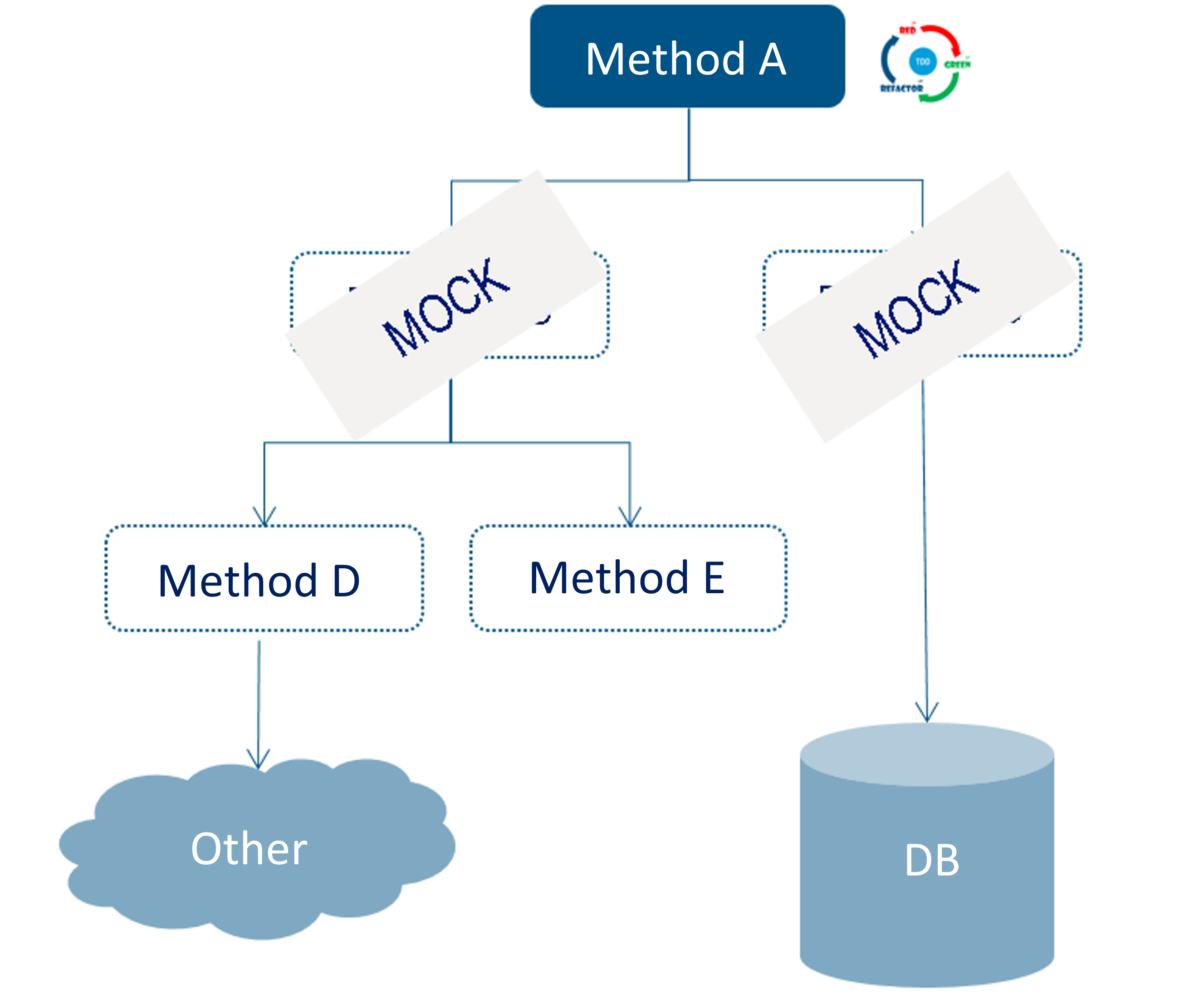}}{Step 2}
     \caption{First steps of the Test-driven Development London School methodology}
     \label{fig:firststepsatdd}
\end{figure}
\FloatBarrier
Once the main method is developed using TDD, the next step is to proceed with the development of the methods on which it depends. These will also be developed using TDD. For each sub-method developed, its substitute can be removed from the main method. In this diagram, the completion of the development of method C in TDD is observed, with its substitute at the level of method A replaced by a real call to method C. This principle will apply to all dependencies as illustrated in steps 3, 4, 5, and 6 of Figure~\ref{fig:stepsatdd}:

\begin{figure}[h]
\footnotesize
\centering
\stackunder[1pt]{\includegraphics[width=6.5cm]{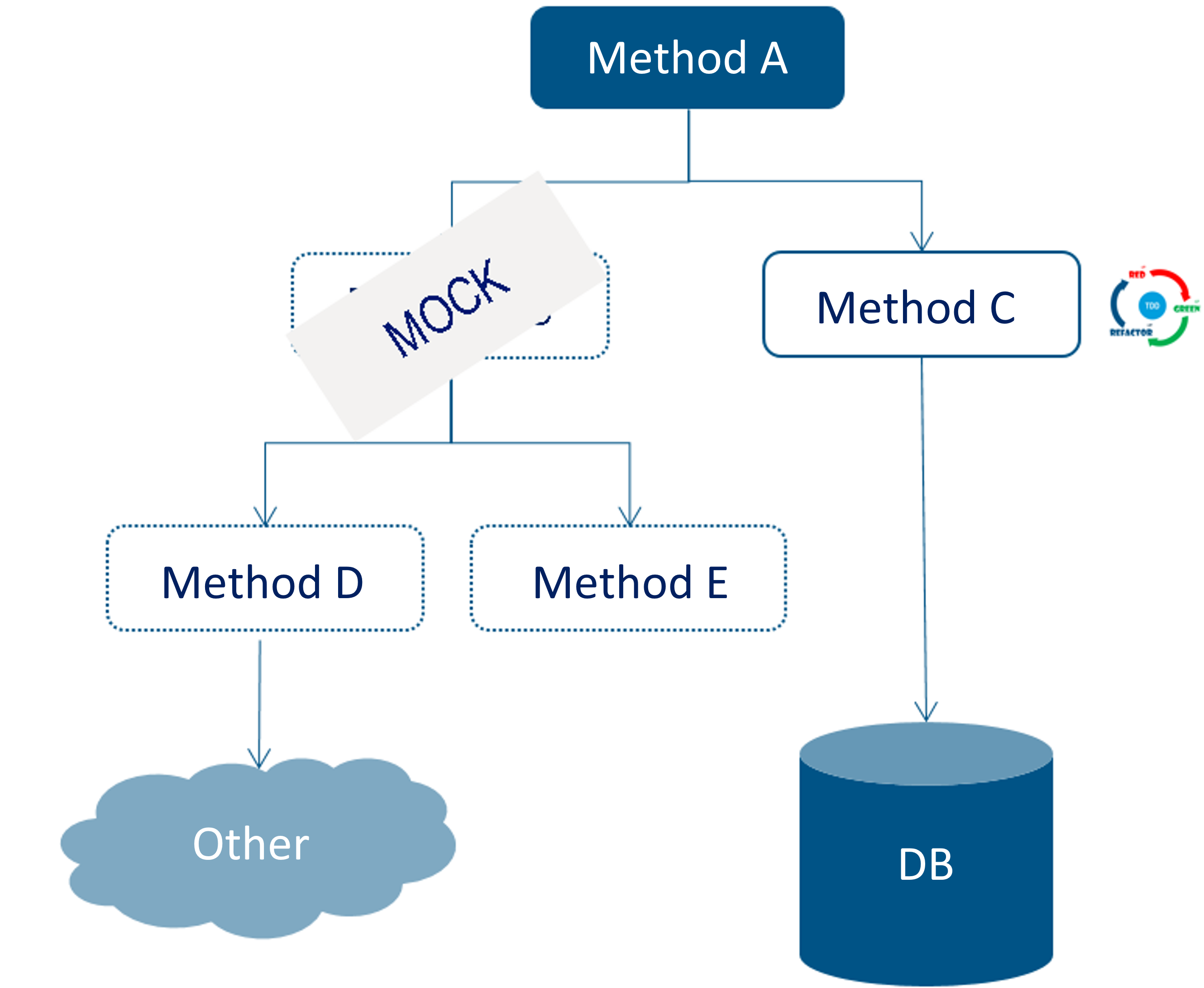}}{Step 3}
\hspace{1cm}
\stackunder[1pt]{\includegraphics[width=6.5cm]{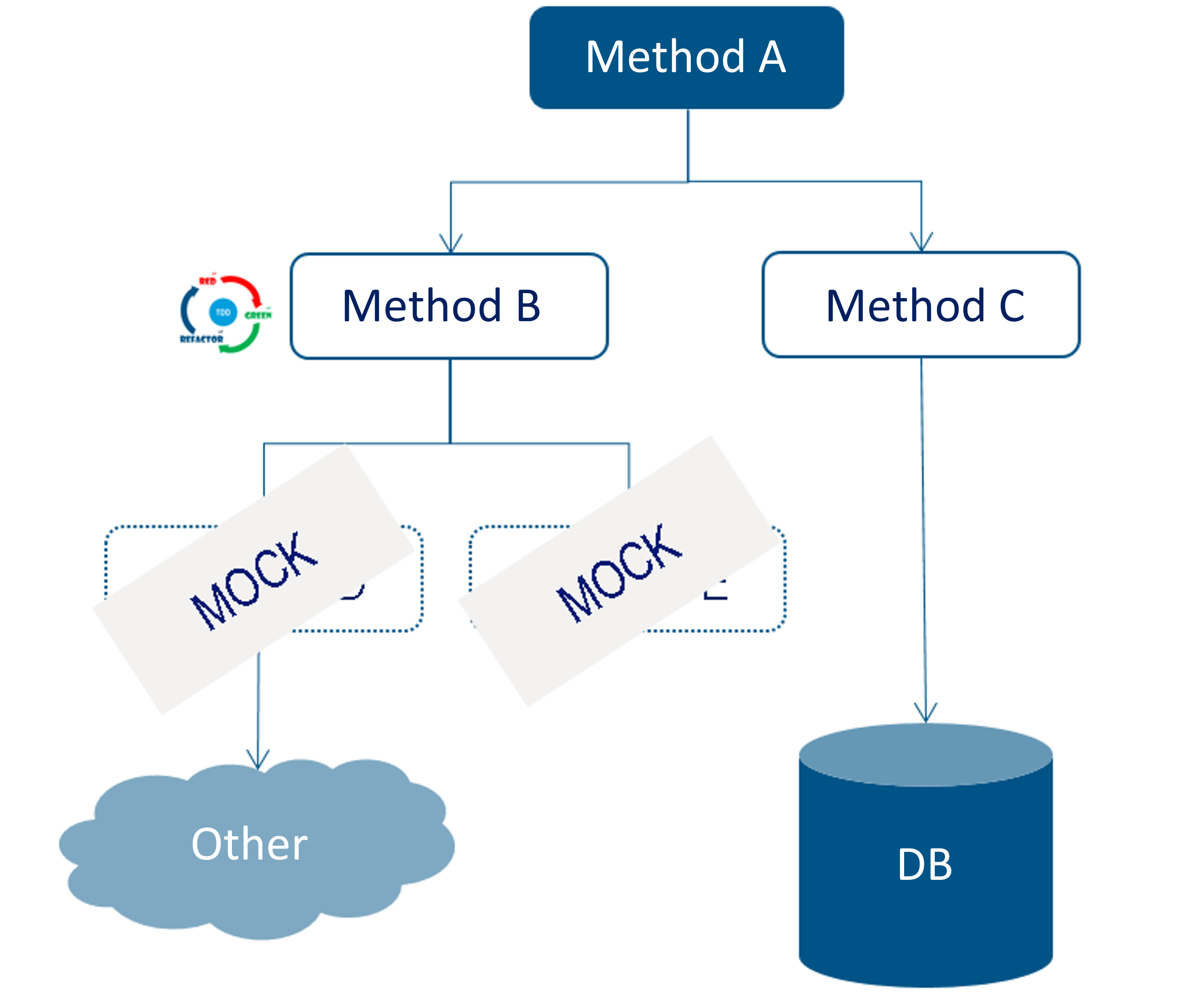}}{Step 4}
\end{figure}

\hfill \break

\FloatBarrier
This approach to TDD allows for breaking down a complex problem into smaller ones while
progressively developing the system as a whole.

\subsubsection{Acceptance Test-driven Development}

ATDD distinguishes itself from TDD "London School" by not using substitutes during
development. Instead, it encourages the developer, when creating method A, to pause
development of this method whenever it requires a functional dependency, and to develop this
dependency separately. This provides the developer with a clear indicator of when their
development is complete: it is the acceptance test of method A. As long as this test is failing
(indicated in red), the developer knows their work is not finished.

Figure~\ref{fig:atdd} illustrates the ATDD Double Loop \citep{Cyrille2022}:

\begin{figure}[h]
	\centering
	\includegraphics[width=.5\textwidth]{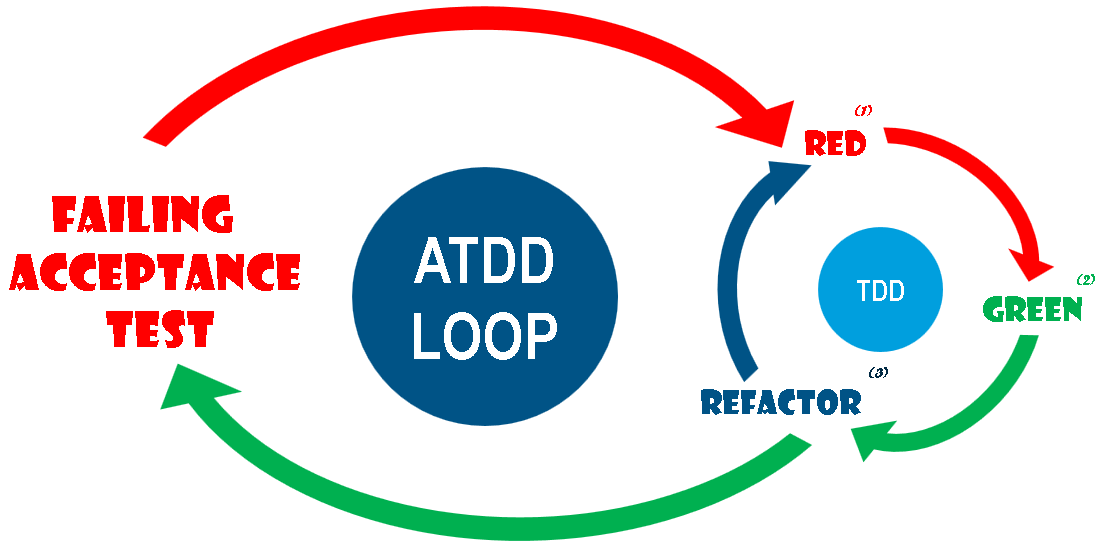}
    \caption{Acceptance test-driven development life cycle}
    \label{fig:atdd}
\end{figure}
\FloatBarrier

ATDD lays the foundation for defining acceptance tests collaboratively, ensuring that stakeholders and developers align on expected outcomes before the development begins \citep{gartner2012atdd}. This collaborative definition of acceptance tests is central to the BDD methodology. As BDD builds upon and extends the principles established by ATDD, particularly in making acceptance tests accessible and formalized in a shared language, it can be seen as an evolution of ATDD.

For this reason, ATDD is not separately evaluated in the comparative analysis, as its core aspects are inherently part of the BDD methodology. The following section describes the BDD methodology in more detail.

\subsection{Behavior-driven Development}

Building on ATDD's focus on acceptance tests, Behavior-driven Development aims to enhance collaboration between stakeholders by introducing a more user-friendly and behavior-focused approach to requirements specification \citep{North2006}.

The first phase of BDD is the “three amigos” meeting. This meeting brings together the three key actors, typically business representatives, QA engineers, and developers, to discuss expected system behavior and clarify acceptance criteria. This phase is important because it combines the knowledge and reasoning of different roles to define expected behaviors in a concrete and shared form \citep{BINAMUNGU2023111749}.

These expected behaviors are often written using Gherkin, a structured, business-readable language used to describe software behavior as executable specifications. Gherkin scenarios are commonly associated with Cucumber, a widely used open-source BDD tool that executes automated acceptance tests from scenarios written in Gherkin. In this context, Cucumber acts as the automation framework, while Gherkin provides the textual syntax used to express the expected behavior. Gherkin relies on keywords that give structure and meaning to executable specifications, including the commonly form \citep{cucumberGherkin2021}:
\begin{itemize}\vspace{-0.5em}
  \item GIVEN \vspace{-0.5em}
  \item WHEN \vspace{-0.5em}
  \item THEN
\end{itemize}\vspace{-0.5em}

By following this structure, the initial state, the desired action, and the expected result are described. This formalism, characteristic of the Cucumber ecosystem, allows Gherkin scenarios to be linked to executable test code through step definitions. These tests are commonly used as acceptance tests because they validate that the produced code conforms to the user stories, i.e., the expected behaviors.

This perspective is closely related to Specification by Example and living documentation approaches, in which concrete examples serve as shared specifications, automated acceptance tests, and continuously updated documentation~\citep{adzic2011specification}.

Those steps are illustrated in Figure~\ref{fig:bdd}:
\begin{figure}[h]
	\centering
	\includegraphics[width=0.8\textwidth]{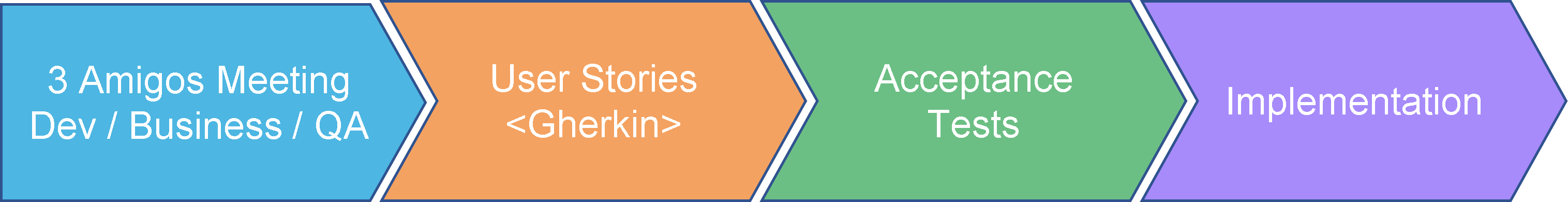}
    \caption{Overview of the Behavior-driven Development workflow}
    \label{fig:bdd}
\end{figure}
\FloatBarrier
BDD offers several key advantages in the context of software development \citep{nascimento2020behavior}. It facilitates strong collaboration among diverse stakeholders, thereby supporting the core values of agility. The formalization of user stories also contributes to maintaining up-to-date documentation throughout the project lifecycle. Furthermore, BDD supports a smooth transition between functional and technical domains by leveraging tools like Cucumber, which bridge the gap between specification and implementation. Lastly, it provides a concrete mechanism for validating the delivered software through automated acceptance tests derived from predefined behavioral scenarios.

While BDD offers clear theoretical advantages for enhancing communication and aligning development with business goals, its practical implementation in large industrial settings often proves challenging. At Dassault Systèmes, for instance, efforts to introduce BDD encountered resistance from non-technical stakeholders, partly due to the mandatory use of Gherkin syntax. Although Gherkin is designed to be close to natural language, its adoption still requires stakeholders across different roles to learn and consistently apply a shared structured formalism, which may differ from their established documentation practices. Business analysts, specification writers and QA teams were frequently reluctant to modify established workflows or adopt a formalism that did not reflect their usual documentation practices. Moreover, the collaborative effort required to write and maintain Gherkin scenarios demands significant time and coordination, which can be difficult to sustain under strict delivery timelines.

These challenges align with findings reported in the literature. Adoption barriers in industrial contexts are frequently organizational rather than technical, particularly when interdepartmental collaboration becomes a bottleneck \citep{irshad2021adapting}. Cultural resistance, limited expertise, and incompatibility with existing processes further complicate BDD adoption at scale \citep{jamesovercoming}. Taken together, these insights suggest that the scalability of BDD may hinge less on its methodological soundness than on its compatibility with the practical and cultural dynamics of industrial software development.

Compared to BDD, ADD enables earlier progress by allowing developers to start designing flowcharts as soon as basic specifications are received. This early design phase often helps clarify ambiguities, similar to how BDD fosters stakeholder collaboration, but with less dependency on ongoing stakeholder input and without requiring non-technical stakeholders to adopt a specific formal syntax or tooling constraints, as algorithmic flowcharts are authored and maintained solely by developers using standard development practices. Flowcharts can be refined iteratively, enabling developers to capture and validate requirements without disrupting the overall workflow. Another notable efficiency gain is the reduced dependency on external stakeholders: whereas BDD often delays implementation until Gherkin-based scenarios are fully defined, ADD allows development to proceed in parallel, refining execution paths independently. In practice, this autonomy accelerates initial implementation while preserving alignment with business objectives.

Furthermore, QA teams are often already familiar with reading logic-based structures such as test flows, condition trees, or decision tables. Similarly, business analysts and specification writers can typically follow flowchart-based logic, especially when diagrams are presented and explained by developers. It is important to note that, in the ADD approach, algorithms are authored and maintained solely by developers. This ensures consistency and technical accuracy while still allowing stakeholders to understand the logic, as it is expressed in natural, non-technical language. This dynamic eliminates the need for stakeholders to adopt a formal syntax, as required in BDD with Gherkin, and facilitates broader acceptance of ADD within existing workflows. In this context, the difference lies not in the intrinsic complexity of the representation, but in the scope of its required adoption across organizational roles.

This realization led to the development of a new approach: an improved version of ATDD, named "Algorithm-driven Development", which will be detailed in Section~\ref{sec:ADD}.

\subsection{Model-Based and Graph-Based Test Derivation}
Model-Based Testing (MBT) is a well-established approach in which abstract behavioral models are used to systematically derive test cases. These models typically represent system behavior using formal or semi-formal structures such as state machines, UML diagrams, or control-flow graphs. Test cases are generated by traversing these models according to defined coverage criteria, enabling systematic exploration of execution paths and improving defect detection. Recent systematic reviews confirm that MBT remains an active research area, with ongoing work focusing on automated test generation, prioritization, and industrial adoption challenges  \citep{mohd2022model} \citep{lonetti2023model}.

Graph-based representations play a central role in MBT and structural testing. Control-flow graphs provide explicit representations of execution paths, enabling systematic derivation of test cases based on structural coverage criteria. Cyclomatic complexity, introduced by McCabe (1976), provides a theoretical foundation for quantifying execution path complexity and determining the minimum number of test cases required for structural coverage. More recent studies have demonstrated the continued use of graph-based models and UML-based behavioral models in MBT for improving test coverage and ensuring systematic validation of software systems \citep{ahmad2019model}.

Despite these advantages, several limitations of traditional MBT approaches have been identified in both academic and industrial contexts. First, MBT often requires the creation and maintenance of explicit behavioral models, which may introduce additional modeling effort and require specialized expertise \citep{mohd2022model} \citep{lonetti2023model}. Industrial experience reports indicate that this modeling overhead can limit adoption, particularly in fast-paced development environments where teams prioritize implementation efficiency and integration with existing workflows \citep{garousi2021model}. Second, MBT models are frequently developed as separate artifacts from the implementation, creating potential inconsistencies between system models and the actual system behavior. Maintaining synchronization between models and implementation can introduce additional complexity and reduce development efficiency. Third, transforming abstract model-based test cases into executable test scripts remains an active research challenge, highlighting practical limitations in bridging the gap between modeling and implementation \citep{ferrari2023transforming}. Finally, traditional MBT approaches focus primarily on structural coverage but do not inherently address redundancy between semantically equivalent execution paths, which may result in inefficient or overly large test suites.

ADD builds upon the fundamental principles of graph-based modeling while addressing these limitations through direct integration into the software development process. Like MBT, ADD represents system behavior using directed graphs in the form of flowcharts, enabling systematic identification of execution paths and derivation of test cases. However, unlike traditional MBT approaches that introduce modeling as a separate activity, ADD integrates algorithmic flowcharts directly into the development workflow as primary artifacts guiding architectural design, test derivation, and implementation. This integration ensures consistency between requirements, architecture, and test generation while eliminating the need for separate modeling phases.

Furthermore, ADD relies on standard flowchart representations that are widely understood by software developers and do not require specialized modeling languages or dedicated modeling tools. This reduces adoption barriers and facilitates practical integration into industrial development environments. In addition, ADD introduces the concept of equivalence blocks to reduce redundancy by grouping semantically equivalent execution paths, enabling systematic yet efficient test derivation. By combining algorithmic modeling, systematic test derivation, and direct integration into the development process, ADD extends the principles of model-based testing into a unified development methodology that improves architectural clarity, test coverage, and development efficiency.

\subsection{Comparative Summary of TDD, BDD, MBT, and ADD}

Table~\ref{tab:comparison_tdd_bdd_mbt_add} positions the rationale of ADD with respect to existing practices. Since ADD is introduced in this paper, the comparison focuses on methodological characteristics such as behavioral modeling, test derivation, and integration into the development workflow, while the method itself is described in detail in Section~\ref{sec:ADD}. Unlike the empirical results presented in Section~4, this comparison is conceptual and aims to clarify how ADD differs from existing approaches.

\begin{table*}[h!]
\centering
\caption{Structural comparison of TDD, BDD, MBT, and ADD methodologies}
\begin{tabular*}{\textwidth}{@{\extracolsep{\fill}} p{3cm} p{3cm} p{3cm} p{3cm} p{3cm}}
\toprule
\textbf{Criteria} & \textbf{TDD} & \textbf{BDD} & \textbf{MBT} & \textbf{ADD} \\
\midrule

Primary artifact 
& Unit tests 
& Gherkin behavioral scenarios 
& Behavioral models (e.g., state machines, UML, CFG) 
& Algorithmic flowcharts \\

Level of validation 
& Unit level 
& Acceptance level 
& Structural/model-based level 
& Acceptance level derived from execution paths \\

Explicit behavior modeling before implementation 
& No 
& Yes (textual scenarios) 
& Yes (formal/semi-formal models) 
& Yes (graphical algorithm) \\

Test derivation mechanism 
& Tests written before code 
& Acceptance tests derived from scenarios 
& Test cases generated from model traversal 
& Test cases derived from flowchart execution paths \\

Modeling phase separation 
& Integrated in coding phase 
& Specification phase before coding 
& Often separate modeling phase 
& Integrated directly into development workflow \\

Requirement for specialized modeling language 
& No 
& Yes (Gherkin) 
& Often yes (UML or formal notation) 
& No (standard flowchart notation) \\

\bottomrule
\end{tabular*}
\label{tab:comparison_tdd_bdd_mbt_add}
\end{table*}

\FloatBarrier

\noindent
The comparison presented in Table~\ref{tab:comparison_tdd_bdd_mbt_add} is based on objective methodological characteristics reported in the literature for TDD, BDD, and MBT \citep{kollanus2011critical, nascimento2020behavior, irshad2021adapting, garousi2021model, ferrari2023transforming}, and on the structural properties of ADD described in this work. 
These criteria reflect observable differences in how each approach represents system behavior, derives tests, and integrates testing activities into the development workflow. 
In contrast, the quantitative evaluation of defect density, test coverage, and development performance presented in Section~\ref{sec:resultsanddiscussion} reports empirical observations on the application of ADD in industrial contexts.

The limitations and practical adoption challenges discussed above motivated the development of ADD in the studied industrial context. The following section presents the resulting methodology, including its core principles, process, and test-derivation mechanism.

\pagebreak
\section{Algorithm-driven Development Methodology}
\label{sec:ADD}
In this software development methodology, a structured approach is proposed to improve alignment between client needs and technical implementation. The methodology takes place after client requirements have been decomposed into distinct tasks, ensuring that each functional unit is clearly defined before development begins.
ADD is designed to promote technical clarity prior to any testing or coding phase. In contrast to conventional workflows that may transition from requirements to development based on implicit assumptions, ADD explicitly formalizes business needs into precise, executable logic through flowcharts. This structured representation reduces ambiguity and aligns technical implementation with functional expectations from the outset.
The process begins with a business document outlining the required scenario. During daily discussions, teams translate these high-level requirements into concrete technical tasks, such as UI elements, backend APIs or other implementation components. Tasks are then assigned to developers, who apply the methodology by creating flowchart algorithms for their respective parts. These diagrams serve as a technical contract, reducing ambiguity and rework while also acting as a visual bridge between client expectations and implementation.

Beyond their role as logical representations, these diagrams define interactions between inputs and outputs, forming the foundation for acceptance tests. These tests are central, as they aim to encompass a broad range of scenarios, including edge cases that may not have been explicitly described by the client. They are then integrated into the code as automated tests, ensuring that the software behaves in accordance with client expectations from the early stages of development.
By offering a structured visual model, this approach reduces the cognitive load on developers, allowing them to focus on refining the logic before diving into complex code. Additionally, the diagrams introduce an architectural dimension by visually organizing responsibilities across components. This is achieved using color coding, which helps distinguish local logic, inter-module calls, and error handling.

This new method is intended to improve the efficiency of the software development process by narrowing the gap between client expectations and implementation, while supporting software quality and reliability. The ADD method consists of three steps:
\begin{enumerate}
    \item \textbf{Writing algorithms using flowchart diagrams:} This step involves modeling the logic of each task through detailed flowcharts. These diagrams aim to explicitly define all relevant execution paths, including edge cases, thereby reducing ambiguity.
    \item \textbf{Extracting the various expected client behaviors from the algorithms:} The algorithms are then analyzed to identify the specific behaviors that the client expects. This helps ensure a comprehensive understanding of the functional requirements before proceeding to the testing phase.
    \item \textbf{Writing acceptance tests for the proper application of the ATDD Double Loop:} Using the identified behaviors, acceptance tests are written to validate the correctness of the implementation through the ATDD Double Loop process, ensuring that the development meets client expectations.
\end{enumerate}

The following diagram~\ref{fig:add_context} illustrates where ADD fits within the agile software development life cycle:
\begin{figure}[h]
	\centering
	\includegraphics[width=1\textwidth]{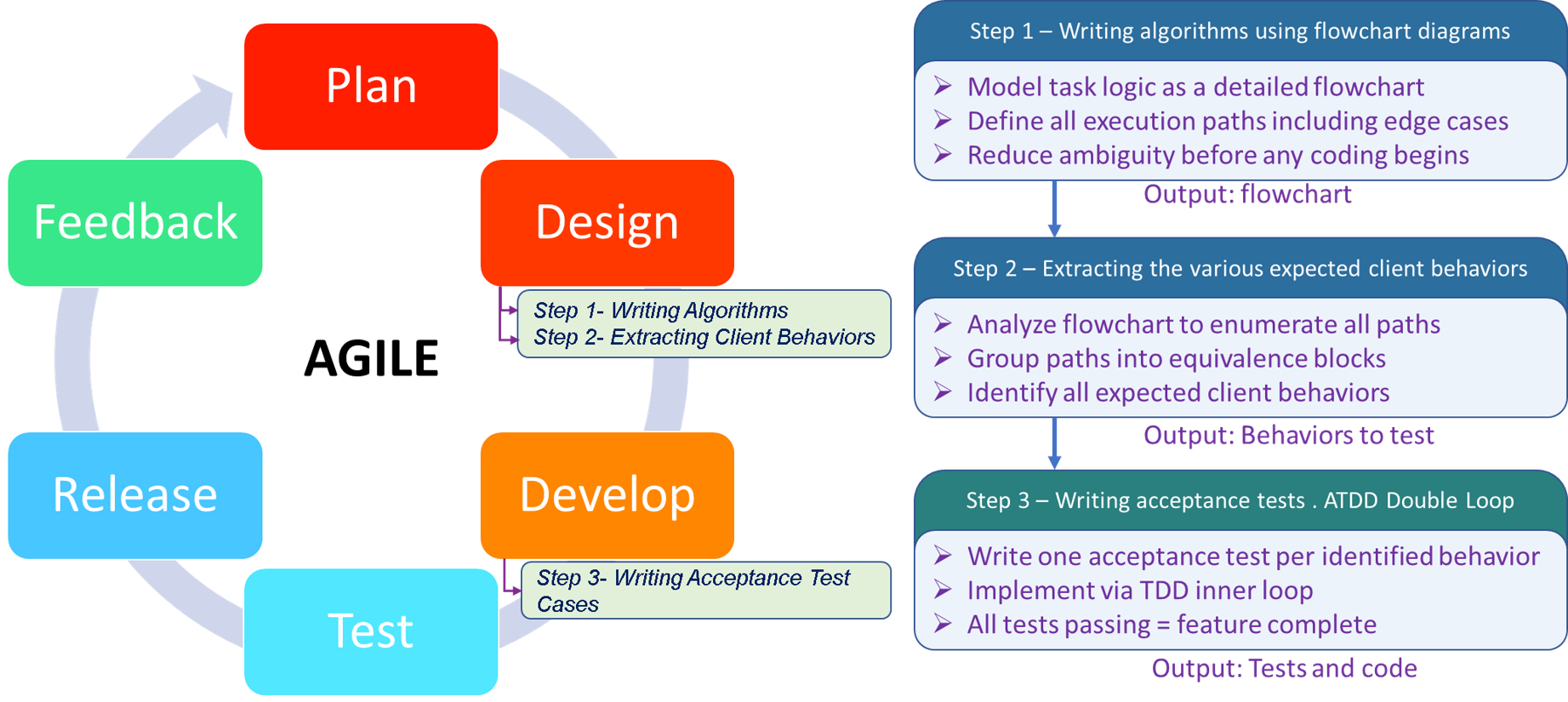}
    \caption{Algorithm-driven Development steps in the agile software development life cycle}
    \label{fig:add_context}
\end{figure}
\FloatBarrier

\subsection{Writing algorithms using flowchart diagrams}

To draft the flowchart diagram effectively, it is essential to first understand the expected client behavior. The first phase involves listening to the client, understanding their needs, and
providing feedback to clarify the expected scenarios effectively.

Once the requirements are understood, it's time for the developer to translate the client's
requirements into algorithms in the form of flowchart diagrams. It is important to note that this translation is performed exclusively by the developer. ADD does not require non-technical stakeholders, such as business analysts or QA engineers, to write or understand algorithms. However, flowcharts may optionally serve as visual aids to support communication when needed or requested by other team members.
The choice of using flowcharts instead of activity diagrams plays an essential role in the systematic extraction of test cases. The reasons behind this choice and its implications for software testing will be discussed in detail in Section 3.2.

\subsubsection{Mathematical Basis of Algorithm-driven Development}

ADD draws upon fundamental concepts of mathematics and logical reasoning to create a structured and formal representation of client requirements. By utilizing algorithmic flowcharts, ADD embodies core mathematical principles such as Boolean logic and decision trees to model the behavior of the system. Boolean logic is used to represent decision points and conditions, ensuring that all possible outcomes are clearly defined. Decision trees visually break down complex scenarios into smaller, manageable branches, which helps in defining and anticipating edge cases comprehensively.

These flowchart diagrams ensure that the software’s behavior is modeled from the outset with rigorous precision, reducing ambiguity and enhancing the accuracy with which client requirements are translated into code.
As will be discussed in Section \ref{sec:add_tests}, this structured mathematical approach also plays an important role in optimizing test case design and ensuring systematic test coverage, reinforcing the rationale for choosing flowcharts over activity diagrams.

In practice, the developer translates these needs into a structured flowchart diagram, specifying both the technical inputs and corresponding steps. A color-coding scheme is typically defined by the development team to indicate the location and nature of the various sub-methods represented in the diagram. The following Table~\ref{tab:color_coding} presents an example of a color-coding convention that can be used in ADD flowcharts:

\begin{table}[h]
\centering
\caption{Example of a color-coding convention used in Algorithm-driven Development flowcharts}
\label{tab:color_coding}
\renewcommand{\arraystretch}{1.8}
\begin{tabularx}{\textwidth}{
  >{\centering\arraybackslash}m{2.5cm} 
  >{\raggedright\arraybackslash}m{13.5cm}}
\toprule
\textbf{Color} & \textbf{Meaning in Flowchart} \\
\midrule
\cellcolor{blue!30}\color{black}\textbf{Blue} &
Indicates that the instruction or sub-method is implemented within the same module. \\
\cellcolor{darkgraycustom}\color{white}\textbf{Gray} &
Indicates that the instruction or sub-method belongs to another module that communicates with an external service. \\
\cellcolor{brightred}\color{white}\textbf{Red} &
Indicates error returns or exception-handling branches. \\
\bottomrule
\end{tabularx}
\end{table}

The flowchart diagram should group instructions that belong to the same level of abstraction. In particular, it should not include the internal logic of methods implemented in other modules. When representing a 'parent' method, the flowchart should display calls to 'child' methods from other modules without expanding their internal logic. If the logic of such a method needs to be detailed, for instance, due to algorithmic complexity, a separate flowchart should be created for that method. It encourages modular design by keeping each flowchart at a consistent abstraction level. This naturally leads to implementing each flow in a single module with one responsibility, following the SRP principle from SOLID~ \citep{martin2018clean}. Cross-module calls are visually identified with a specific color code, which also helps maintain clear separation of concerns. Although ADD does not enforce object-oriented modularity (i.e., decomposition at the class level), it remains fully compatible with it. By reducing the abstraction level of flowcharts from modules to classes, the resulting design can lead to multiple diagrams, each reflecting an individual class responsibility. This adjustment enables ADD to support object-oriented implementations where algorithms are grouped into cohesive class units aligned with SRP, thereby bridging modular development at both the architectural and class levels. Naturally, object-oriented features such as inheritance, polymorphism, and interface-based design can be incorporated at the implementation level, depending on the development context, but are not mandated by the ADD methodology itself.

When drafting the flowchart diagram, the developer can refine their understanding of the requirements and anticipate behaviors not initially described by the client. This includes identifying additional cases, such as verifying user permissions, validating resource existence, or handling unexpected input conditions. Anticipating such scenarios helps to reduce incident reports during development and contributes to overall system robustness.

Once the algorithm is translated into a flowchart diagram, the developer may choose to contact the client again to validate newly identified scenarios, if necessary. Additionally, the algorithm should be shared with the QA teams to facilitate the writing of test cases and ensure alignment between design and verification activities.

Beyond its value as a communication tool, the flowchart diagram also serves as an architectural plan. It allows developers to visualize the complexity of the scenario, to identify the modular structure of the code, and to anticipate potential concerns related to performance, capacity, and scalability.

These three aspects are essential for developers as they help distinguish proficient developers from less experienced ones. The key competence lies in the ability to effectively manage algorithmic
complexity, meet customer needs, and proactively anticipate performance, capacity, and
scalability issues in the application. This triple competency is more impactful than extensive knowledge of specific programming frameworks or having a perfect understanding of clean code principles.

On the other hand, these flowcharts will be extremely useful for new developers joining the team, as they will allow them to quickly gain expertise in the application's domain and code architecture. This will prevent them from having to delve into thousands of lines of code without being able to establish a connection with the domain, which could lead to confusion or frustration. However, this benefit is contingent on a key boundary condition: the effectiveness of ADD relies on the completeness and accuracy of the algorithmic diagrams. If relevant edge cases or failure scenarios are not identified and represented during the diagramming phase, they will neither be designed for nor tested. This limitation is inherent to diagram-driven test derivation.

Having established the role of flowchart-based algorithms in representing execution logic and supporting early design decisions, the next step is to derive expected client behaviors from these algorithms.

\subsection{Extraction of the various expected client behaviors from the algorithms}
\label{sec:add_tests}

Understanding client behavior is essential, but ensuring that this behavior is correctly implemented and tested is even more critical. One of the core strengths of ADD lies in its ability to design high-quality tests by effectively working through the algorithmic flowchart. The challenge lies in achieving comprehensive and meaningful coverage of all client scenarios modeled within the flowchart diagrams.

ADD aims to systematically identify and test all execution paths within a diagram, including edge cases, from the early stages of development. A key element of this methodology is the use of algorithmic diagrams to visually structure execution logic, making it easier to derive test cases that reflect expected behaviors.

For effective test derivation, developers should be able to determine the number of execution paths, and consequently the number of test cases, at a glance. As stated by McCabe \citep{wijendra2021analysis}, the cyclomatic complexity metric quantifies the number of linearly independent execution paths within a flowchart diagram. Since each flowchart path represents a unique execution route, its cyclomatic complexity directly correlates with the minimum number of test cases required. This makes flowcharts not only a powerful visualization tool but also a structured approach for ensuring comprehensive test coverage and minimizing the risk of overlooking critical execution paths.

While cyclomatic complexity provides a valuable measure, it only gives a lower bound on the number of test cases required to cover all decision points. This is why ADD focuses on ensuring that all control-flow decisions (branches and conditions) are exercised at least once, regardless of the number of paths.

Another key advantage of flowcharts in ADD is their alignment with modular coding practices, particularly in multithreaded systems, where execution logic is divided into independent functional units. When constructing flowcharts, it becomes intuitive to separate each thread into distinct diagrams, enhancing readability, modularity, and maintainability. This structured decomposition ensures that each module remains independent, facilitating scalability and test automation.

While flowcharts are the preferred tool for ADD, activity diagrams still have their place, particularly for representing high-level process interactions and business logic.

The following section presents a systematic method to extract test cases from flowcharts while accounting for algorithmic complexity and preparing the ground for later optimization through semantic equivalence.

\subsubsection{Systematic Test Case Extraction Using Flowcharts}

The structured representation provided by flowcharts makes test case extraction in ADD systematic and traceable. Each path represents a complete scenario that the system may encounter. By following every execution route between inputs and outputs, developers can identify the structurally distinct scenarios and translate them into acceptance tests, including relevant edge cases. The quality of these tests depends directly on the precision of the algorithm design. A well-designed algorithm explicitly models each branch, decision point, and flow, thereby providing a basis for deriving logical and comprehensive test cases.

In traditional approaches, test design is often an ad hoc process in which developers consider edge cases and dependencies after writing the initial code. With ADD, the flowchart guides test design proactively from the outset, making tests a core part of the system blueprint rather than an afterthought. Each modeled branch can be translated into a test, helping ensure that the expected behaviors are represented. When the algorithm has been validated and reviewed, the resulting tests remain aligned with the requirements and client needs. The visual representation also helps developers structure tests effectively while reducing the risk of overlooking edge cases or introducing redundant tests.

Transforming the algorithm into tests supports coverage by providing a clear mapping between requirements and system behaviors. This is an important developer competency because software quality depends strongly on test design. In ADD, systematically derived tests are intended to cover the required functionality while remaining maintainable and extensible as the project evolves.

However, the number of acceptance tests can grow rapidly with algorithmic complexity. This is evident in a basic algorithm composed of two conditional blocks, each having two possible outcomes. Two successive conditions therefore produce $2 \times 2$ possible combinations, resulting in four potential client scenarios.

Before examining this issue further, two basic rules should be established. First, the algorithm should represent only one level of abstraction at a time. Second, both outcomes of each conditional block should be tested.

This generation step produces a complete initial set of tests. In practice, this set can often be reduced without loss of relevant coverage, as explained in the next section.

\subsubsection{Test Case Optimization and Equivalence Blocks}

After generating the initial exhaustive set of test cases, ADD introduces heuristics for reducing redundancy while maintaining relevant coverage. The goal is to balance test completeness and practicality, avoiding the pitfalls of combinatorial explosion.

As mentioned earlier, cyclomatic complexity provides only a lower bound on the number of test cases. This limitation becomes particularly evident in the presence of cycles or loops, where the number of possible execution paths may become unbounded. In such cases, the goal shifts from exhaustive path enumeration to achieving meaningful coverage based on control-flow semantics. ADD focuses on ensuring that loop structures are tested under representative conditions (e.g., zero, one, and multiple iterations). The determination of how many loop iterations to test is left to the developer’s judgment, based on system requirements and domain knowledge.

At the design level, developers should also ensure that two conditional blocks cannot be merged into one. If possible, merge them to reduce unnecessary branching.

As the number of independent conditions grows, the total number of execution paths can increase exponentially. However, not all combinations yield semantically distinct behaviors. Some paths, though structurally different, may lead to identical downstream operations or system states.

To address this complexity, the concept of \textit{equivalence blocks} is introduced. These are parts of the algorithm for which the differences between the paths taken inside the block do not affect the evaluation or the effects of subsequent conditions (for example, an \textit{if} statement). In such cases, only a single representative path needs to be tested, provided that all individual conditions are exercised elsewhere to ensure full decision coverage~\citep{sasmito2023ecp}. This approach allows redundant acceptance tests to be eliminated without compromising system quality.

This heuristic shares foundational motivation with classical techniques such as Equivalence Class Partitioning and Modified Condition/Decision Coverage (MC/DC)~\citep{hong2020modified}. However, it operates at a higher semantic level: while MC/DC focuses on isolating the effect of individual boolean conditions within expressions, equivalence blocks assess independence and impact across entire execution paths. Their identification is guided by the flowchart and the developer’s understanding of control-flow semantics and domain knowledge.
To ensure clarity and reproducibility, the concept of equivalence blocks can be formally defined as follows.

\pagebreak
\medskip
\noindent\textbf{Formal Definition of Equivalence Blocks.}

Let an algorithm be represented as a directed graph $G = (N, E)$, where $N$ is the set of nodes representing operations or decision points, and $E$ is the set of directed edges representing control flow. A path $p$ is defined as a sequence of nodes from the entry node to an exit node.

An equivalence block $B \subseteq N$ is defined as a subgraph of $G$ such that, for any two execution paths $p_1$ and $p_2$ that differ only within $B$, the observable system behavior after exiting $B$ remains identical. Observable behavior includes outputs, state transitions, database operations, and interactions with external modules or dependencies.

Two paths $p_1$ and $p_2$ are considered equivalent with respect to $B$ if and only if their execution results in identical observable behavior after exiting $B$. In such cases, a single representative path is sufficient to validate the behavior of all paths belonging to the same equivalence block.

This definition allows the test set to be reduced while preserving behavioral coverage, ensuring that each distinct observable behavior is exercised by at least one test case.

\medskip
\noindent\textbf{Formal Path Coverage Heuristic Based on Equivalence Blocks.}

Given a flowchart graph $G = (N, E)$, the ADD test generation heuristic proceeds as follows:

\begin{enumerate}

\item Identify all decision nodes and possible execution paths.

\item Partition execution paths into equivalence blocks based on identical observable behavior after exiting the block.

\item Select at least one representative path for each equivalence block.

\item Ensure that each decision outcome and loop structure is exercised by at least one selected path.

\end{enumerate}

This heuristic ensures systematic coverage of distinct observable behaviors while avoiding redundant test cases, providing a scalable approximation of exhaustive path coverage.
Each selected representative path is then translated into a corresponding acceptance test case, ensuring traceability between the algorithmic model and the resulting test suite.

\medskip

\noindent\textbf{Example: Lambda Algorithm}

The following example demonstrates how identifying an equivalence block can reduce the number of acceptance tests from 12 to 9 without loss of coverage. Consider the following simplified lambda algorithm represented in Figure~\ref{fig:lambda}:

\begin{figure}[h]
	\centering
	\includegraphics[width=0.4\textwidth]{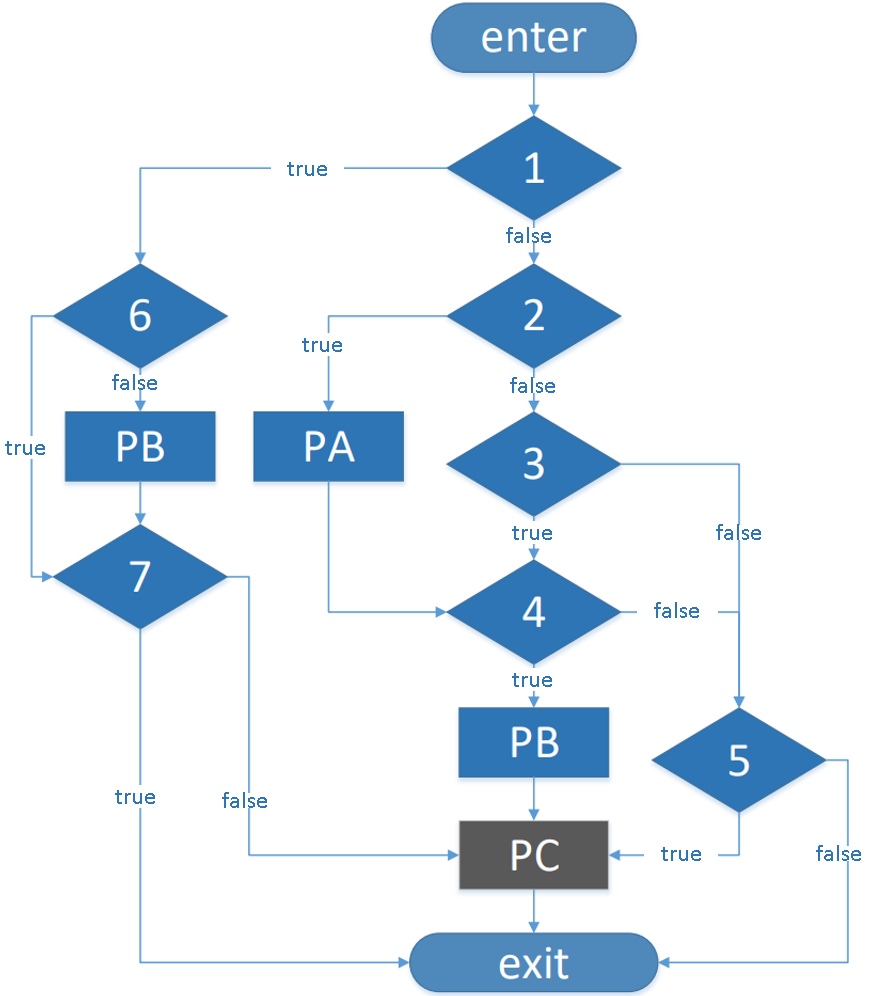}
    \caption{Lambda algorithm diagram}
    \label{fig:lambda}
\end{figure}
\FloatBarrier
The numbers in the algorithm represent conditional blocks, while the instruction blocks are
indicated by letters "Px" (for 'process'). 
Representing all possible paths yields 12 combinations
 (Figure~\ref{fig:combinations}):

\begin{figure}[h]
	\centering
	\includegraphics[width=0.90\textwidth]{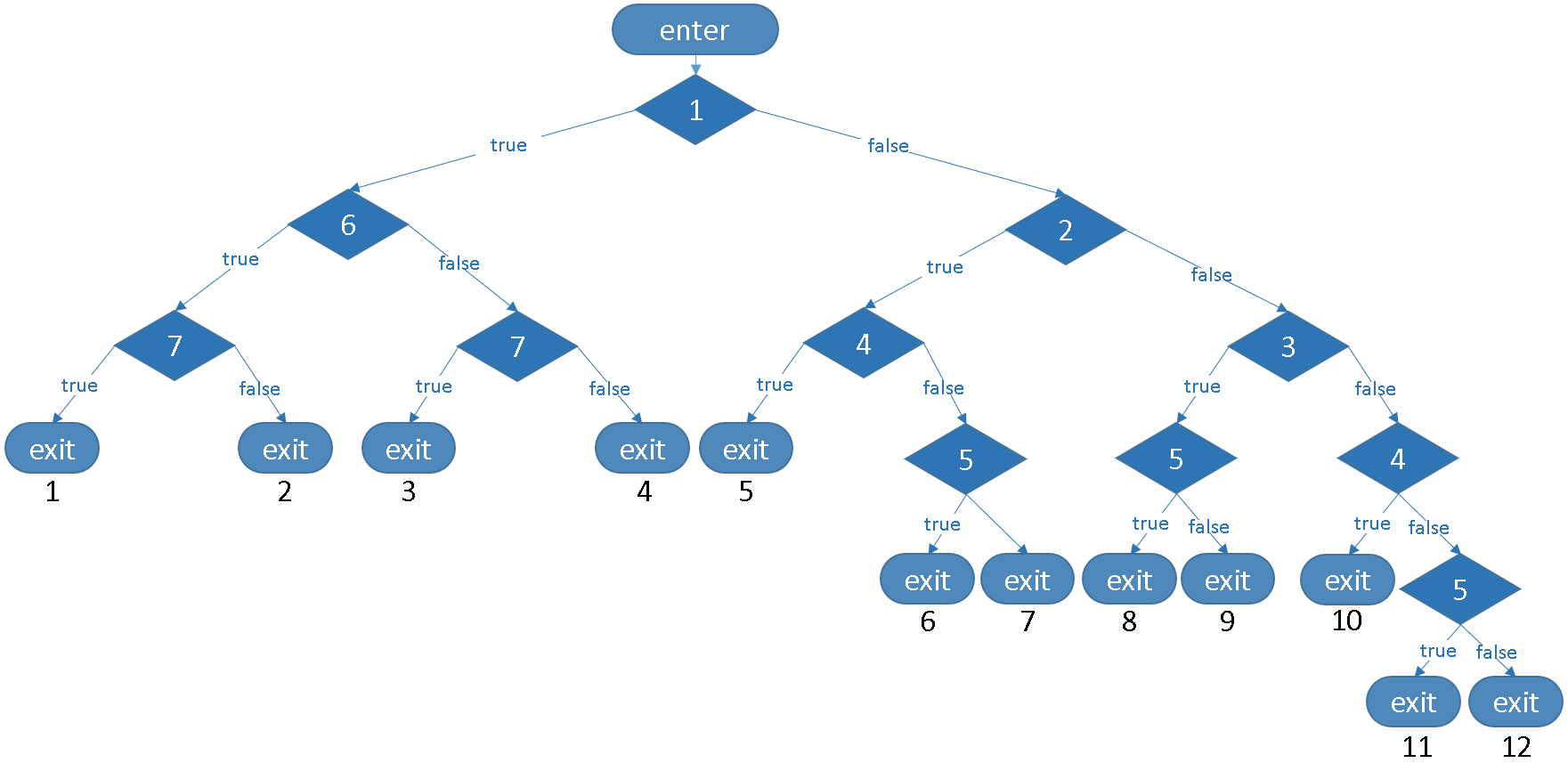}
    \caption{Visualization of All Potential Combinations in the Algorithm}
    \label{fig:combinations}
\end{figure}
\FloatBarrier
As seen even for a relatively simple algorithm, the number of acceptance tests can be
high.

It is therefore relevant to consider whether some acceptance tests may be eliminated if they do not meaningfully contribute to system stability or the fulfillment of client requirements. In many cases, such reductions are possible and desirable to optimize the time allocated to testing activities. By carefully examining the branches of the algorithm (the paths between two blocks), it is possible to determine if each branch depends on the conditions preceding it. In the diagram, the process "PC" is deliberately colored differently to indicate that this process depends on the different branches taken. Consider a concrete example: an object 'dog' stored in the database at process "PC". This object can be modified upstream by process "PA". Therefore, the actual behavior of "PC" depends on the path taken before its execution.

On the other hand, all branches preceding condition "5" are independent. This means that these
branches form an equivalence block. Figure~\ref{fig:equivalentblock} illustrates the algorithm with the equivalence block enclosed:

\begin{figure}[h]
	\centering
	\includegraphics[width=0.4\textwidth]{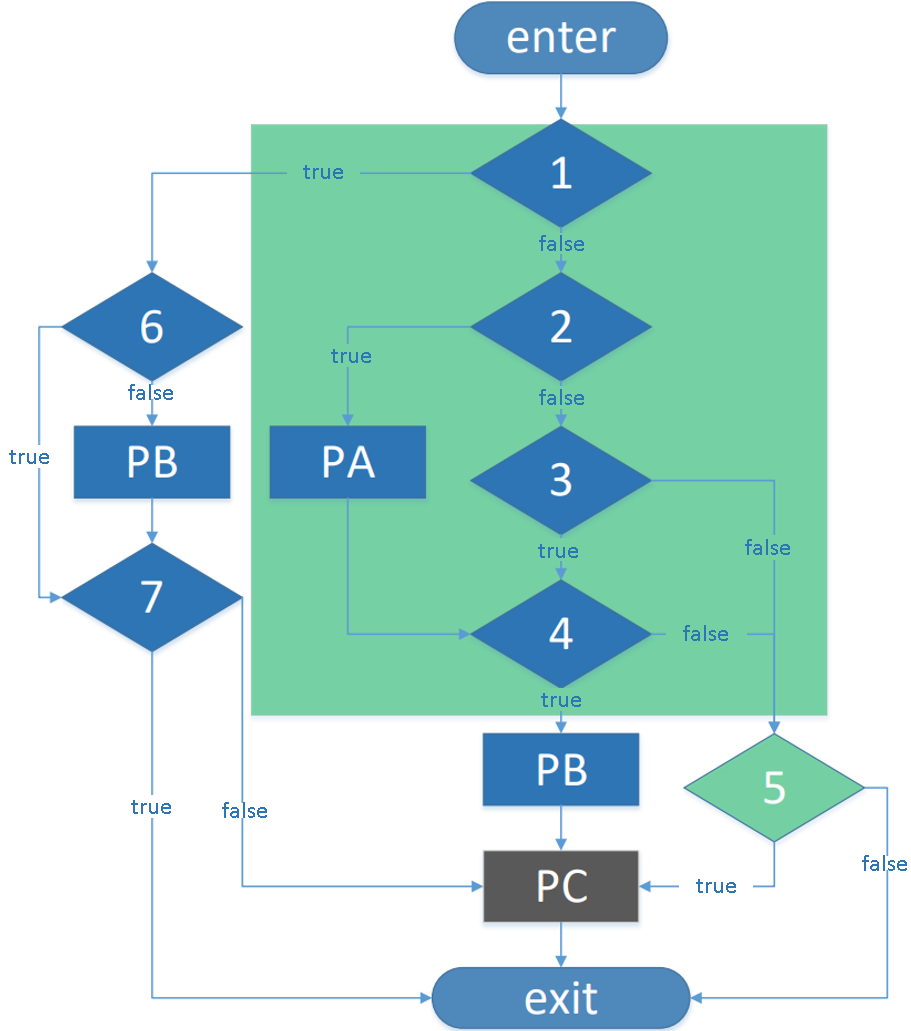}
    \caption{Equivalence blocks are highlighted in green.}
    \label{fig:equivalentblock}
\end{figure}
\FloatBarrier

More precisely, regarding condition 5, the path taken does not influence the outcome of this
condition. As mentioned earlier, while it is necessary to test both possible outcomes of each
conditional block, it is not always necessary to test all combinations of paths leading to this
condition. By applying these principles, redundant acceptance tests can be eliminated without compromising coverage. In this example, three scenarios are skipped, as shown in Figure~\ref{fig:combinations_filtered}:

\begin{figure}[h]
	\centering
	\includegraphics[width=0.90\textwidth]{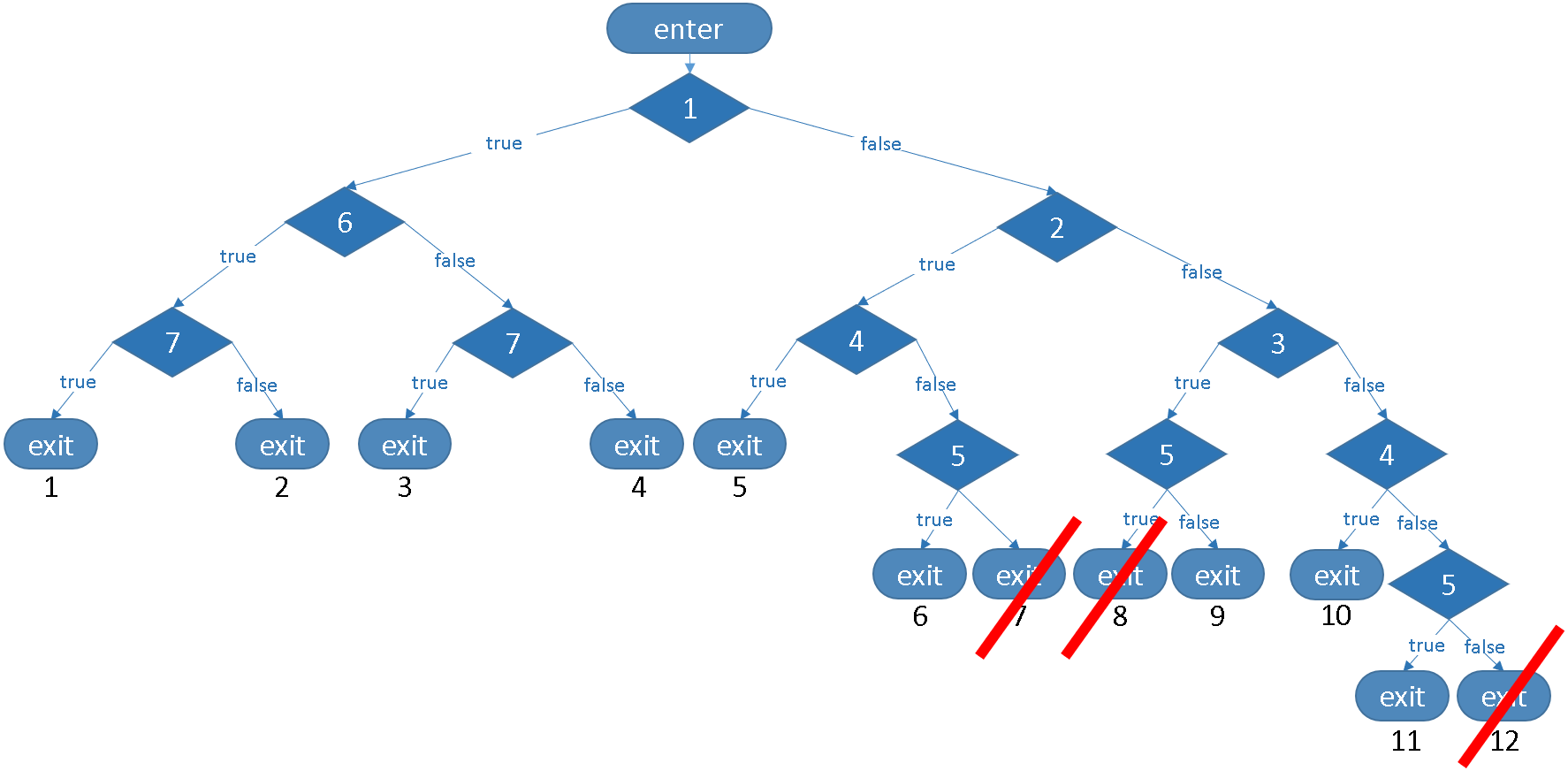}
    \caption{Visualization of All Potential Combinations to Be Skipped}
    \label{fig:combinations_filtered}
\end{figure}
\FloatBarrier

\subsection{Writing acceptance tests for the proper implementation of the ATDD Double Loop}

The ADD method integrates the ATDD Double Loop into its third step. As in TDD, this phase begins by writing tests before implementation. According to the principles of TDD, only one test should be written at a time. It is recommended to start with the shortest paths to develop in a linear and successive manner. By following this method, the developer minimizes the need to revisit previously developed code lines.

Once the acceptance test is written, the ATDD Double Loop method should be applied to structure the development process effectively. This approach will enable the developer to know when their development is complete. This process encourages incremental development, progressively covering anticipated and specified use cases in alignment with client expectations.

These tests provide an essential guarantee: they ensure not only that the code meets the clients'
needs, but they also serve as protection against potential regressions when integrating new
developments by the developers. This approach enhances the reliability of the software
throughout its development cycle.

Every future evolution should be integrated first into the flowchart before drafting its new
acceptance test.

Most defects observed in production can be traced to one of the following sources:
\begin{itemize}
    \item The developer misinterpreted the client's needs in their flowchart.
    \item The developer incorrectly wrote their acceptance test.
\end{itemize}

In both cases, the developer must verify that their algorithm supports the scenario described in
the incident report. If not, they should integrate this scenario into their flowchart before
proceeding to write the acceptance test.

The margin of error is reduced, as it resides in a small part of the development
cycle rather than spread over thousands of lines of production code.

It is suggested that each developer undergoes a cross-review of flowcharts and acceptance tests with colleagues. In the studied context, developers reported that this mutual review process was more focused and easier to conduct than traditional code reviews, which often require navigating large portions of production code. By relying on visual flowcharts and their corresponding acceptance tests, reviewers can more easily identify inconsistencies between the intended behavior, the modeled execution paths, and the implemented test cases. This supports earlier detection and correction of specification, design, or test-related errors.

In agile and Extreme Programming contexts, rapid feedback and continuous improvement are central principles, but review activities must remain compatible with the time constraints of iterative delivery \citep{Beck1999, akhtar2022extreme, cao2004extreme}. Therefore, review time should be used efficiently to detect errors early without compromising development cadence.

\subsection{Writing Quality Tests in Practice}

In this paper, the term \emph{test case} is used as a generic term for any executable validation scenario, including unit-level tests and higher-level functional tests. An \emph{acceptance test} refers more specifically to a test case that validates a complete expected behavior of a feature or API from the user's or client's perspective. In ADD, acceptance tests are derived from flowchart execution paths and are used to verify that the implemented functionality conforms to the expected behavior.

Within the ADD methodology, the process of writing high-quality tests is directly driven by the algorithmic flowchart. Once a test path is selected, the corresponding test must validate both the expected output and the internal behavior defined by the algorithm. This includes verifying not only the final result of the execution path but also all intermediate operations, especially those involving interactions with external modules, frameworks, or services.

The intrinsic quality of written tests directly impacts the overall effectiveness of the software validation process~\citep{peng2021unit}. In software testing, test quality is commonly characterized by several established dimensions, including coverage (the extent to which execution paths, branches, or conditions are exercised by the test suite), fault detection effectiveness (the ability of tests to detect defects when present), isolation (ensuring that the behavior under test is evaluated independently of unrelated modules), and maintainability (ensuring that tests remain understandable, traceable, and adaptable as the system evolves)~\citep{ammann2017introduction}. Within the ADD methodology, these properties emerge from the systematic derivation of test cases from the algorithmic flowchart, which explicitly defines execution paths and system behavior.

To implement this, the developer begins by identifying the sequence of operations and decision points along the selected path in the flowchart. Each operation in this sequence is then translated into a corresponding segment of test logic. When the flowchart indicates a conditional branch that should lead to an exception, the test must explicitly assert that the correct exception is raised along with the appropriate error message. These assertions are typically implemented using constructs such as \texttt{assertThrows} in JUnit~\citep{junit} or equivalent features in other modern testing frameworks.

When the flowchart indicates that certain operations invoke external methods or modules, often represented through specific color conventions, these interactions must be mocked and verified. In Java, for instance, frameworks such as Mockito~\citep{mockito} allow developers to simulate and inspect such calls. The \texttt{verify} method, for example, can assert that a specific dependency was invoked with the expected parameters. This step ensures that the structural expectations defined by the algorithm are properly enforced.

For each test, it is crucial to isolate the logic under test. Dependencies must be mocked to prevent side effects and ensure test determinism. The flowchart also defines the scope of what should be tested in isolation. Any process box that belongs to another module, as visually indicated by color or labeling, is generally mocked rather than tested directly. This separation helps preserve modularity and aligns with the ADD philosophy of clearly localizing responsibilities.

The completeness of a test can be assessed by its ability to exercise the intended execution path and validate the expected system behavior, including both expected outcomes and absence of unintended interactions. For example, if a particular flow is not supposed to trigger a database write, the test may assert the absence of such a call, using constructs such as \texttt{verify(..., never())} in Mockito. Such assertions help ensure that the system behaves according to its specified logic.

Additionally, the naming and internal structure of each test should directly reflect the logic of the flow it covers. Method names such as "\texttt{methodName\_conditions\_outcome()}" provide immediate clarity about the scenario being tested and the expected outcome. This contributes to both readability and long-term maintainability, especially in larger test suites.

The algorithmic flowchart informs not only the assertions to be made but also the identification and configuration of mocks and dependencies. This ensures consistency between the visual design and the implemented behavior. Unlike approaches that rely solely on interpreting textual requirements, ADD provides an explicit behavioral model from which test cases are systematically derived.

In summary, writing quality tests with ADD consists in methodically transforming each part of the flowchart, its conditions, operations, and module boundaries into structured test logic. The quality of the resulting test suite emerges from its systematic coverage of execution paths, its ability to validate system behavior, and its maintainability over time.

After defining the ADD methodology, the following section discusses how it can be integrated into contemporary software engineering workflows, particularly Agile and DevOps environments.

\subsection{Integration of ADD into Modern Software Practices}
ADD was designed with compatibility in mind for contemporary software engineering workflows. To evaluate its practical adoption, this section explores how ADD integrates with two dominant paradigms in modern development environments: Agile methodologies and DevOps practices. These perspectives provide insight into how ADD complements iterative planning, continuous delivery, and automation workflows without disrupting existing team dynamics or technical pipelines.

\subsubsection{ADD in Agile Software Development}
The ADD methodology is naturally aligned with Agile principles, as it mitigates bottlenecks before coding begins. Much like Agile, ADD promotes continuous feedback and collaboration, as flowcharting forces teams to clarify requirements early. By structuring algorithms before development starts, teams can significantly reduce rework and avoid costly late-stage modifications.

Furthermore, ADD improves development predictability. Since each algorithm represents a self-contained unit of work, teams can accurately estimate timelines based on the complexity of the defined logic. This structured approach prevents unexpected scope creep, ensuring that Agile sprints remain focused and manageable.

Additionally, ADD is inherently iterative, aligning with Agile’s incremental delivery model. Each development cycle follows a structured flowchart, allowing teams to integrate algorithmic design into backlog refinement and sprint planning. The confidence gained from ADD enables teams to implement frequent changes during sprints without breaking existing functionality.

Finally, by reducing the number of late-stage defects, ADD contributes to maintaining a sustainable and constant delivery pace, consistent with Agile principles of sustainable development.

\subsubsection{ADD and DevOps Integration}
\begin{figure}[h]
	\centering
	\includegraphics[width=0.4\textwidth]{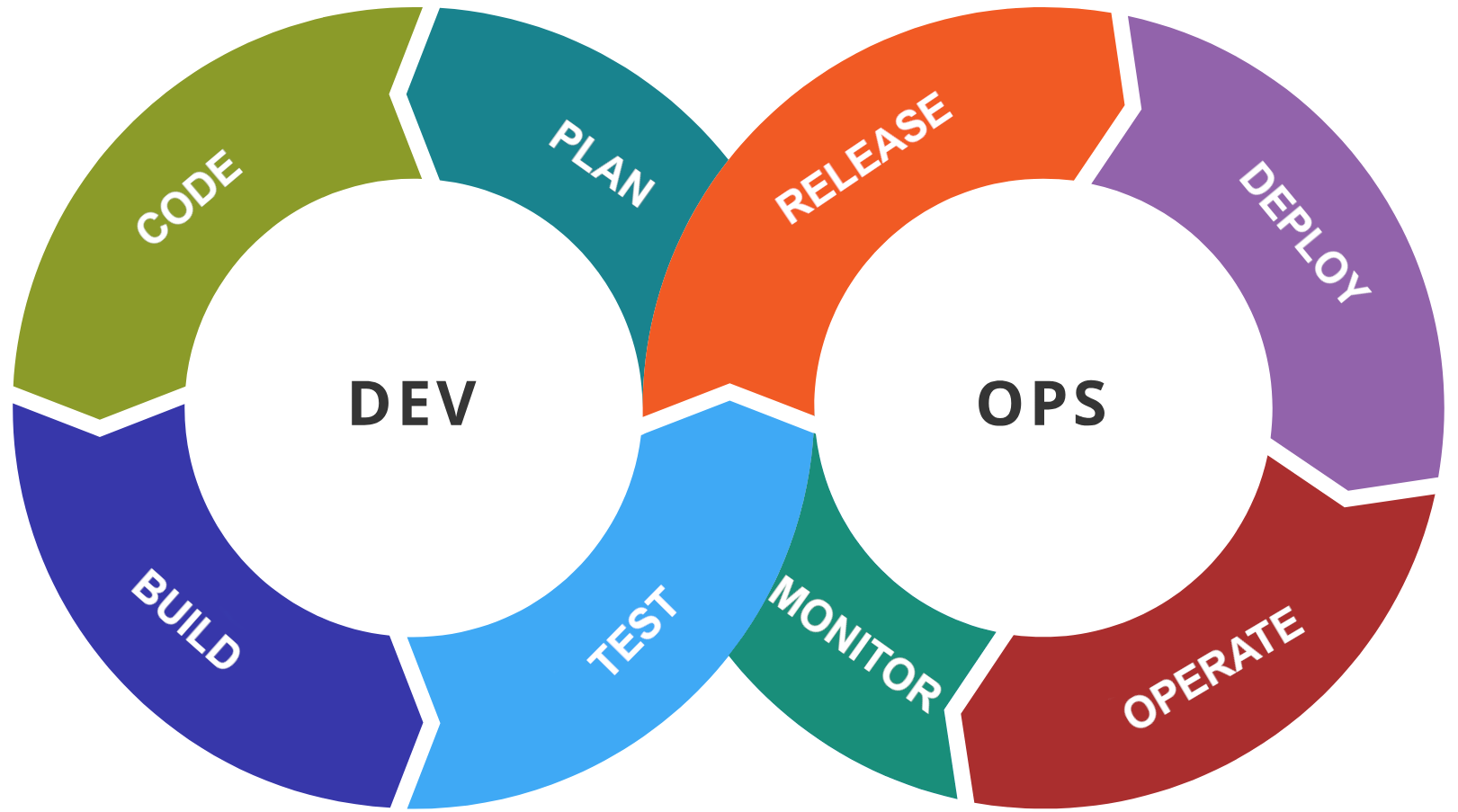}
     \caption{DevOps Life cycle}
     \label{fig:devops}
\end{figure}

Beyond Agile, ADD seamlessly integrates into the DevOps life cycle, reinforcing best practices across each phase of CI/CD. By formalizing execution paths upfront, ADD enhances pipeline stability and deployment reliability, addressing five critical stages of the DevOps cycle illustrated in Figure~\ref{fig:devops}.

\FloatBarrier

During the planning phase, algorithm definitions contribute to reducing ambiguity in feature development. In the coding stage, the use of structured logic leads to cleaner and more maintainable implementations. When building, well-defined flowcharts help prevent integration issues and resolve dependency conflicts. In the testing phase, ADD supports early and automated testing, thereby improving both test coverage and overall system stability. Finally, during the release phase, the approach ensures that only stable, high-quality code is introduced into the deployment pipeline.

These contributions align with empirical evidence on high-performing DevOps practices, which highlight early testing, continuous automation, and reliable deployment pipelines as key drivers of software delivery performance \citep{forsgren2018accelerate}. Furthermore, by embedding specification logic directly into the development process, ADD operationalizes architectural principles that support maintainability, reliability, and resilience, concerns central to DevOps from an architectural perspective \citep{bass2015devops}. In this way, ADD provides a structured mechanism for translating requirements into executable and verifiable logic, strengthening both technical quality and operational outcomes in DevOps environments.

\section{Results and Discussion}
\label{sec:resultsanddiscussion}

This section evaluates the practical application of ADD through two empirical studies conducted with different software development teams. The first involves Team 1 and examines the methodology’s operationalization across three progressively complex scenarios, ranging from a basic technical function to a critical, large-scale migration. These examples demonstrate ADD’s applicability at both the micro level (technical design) and macro level (project structuring). A detailed analysis of the team’s performance follows, supported by productivity and quality metrics collected over four years of sustained use.

The second study focuses on Team 2, which adopted ADD incrementally during the redesign of a core system. In contrast to Team 1, this team transitioned from conventional development practices, offering insight into the method’s adaptability within an existing engineering culture. Both quantitative indicators and qualitative observations are presented, particularly regarding defect rates, complexity management, and team coordination.

The results presented in this section are grounded in industrial practice and reflect the application of ADD in operational contexts. While these case studies do not aim to provide statistically generalizable proof, they offer insight into how the methodology performs under real-world delivery constraints, with varying levels of team maturity, project complexity, and organizational expectations.

The results presented here aim to assess the methodology’s viability, effectiveness, and scalability under diverse conditions of integration, functional complexity, and organizational context. The section concludes with a discussion of the limitations and boundary conditions observed during these implementations.

\subsection{Comprehensive Evaluation of ADD within Team 1}
\label{sec:team1}
\begin{figure}[h]
\centering
\includegraphics[width=1\textwidth]{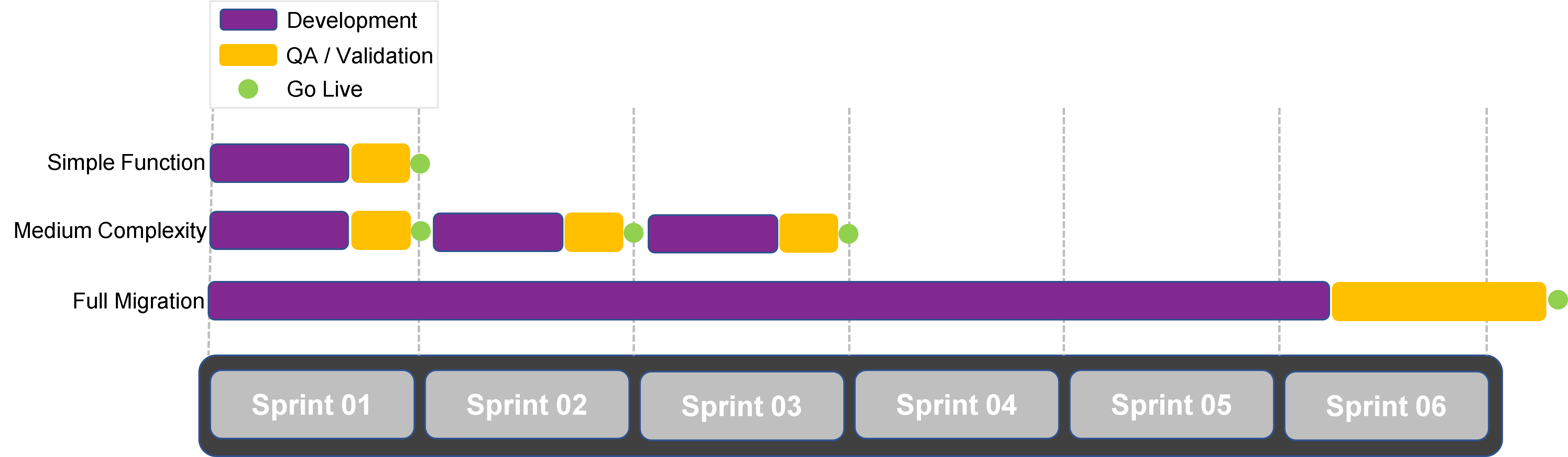}
\caption{Overview of the three examples illustrating Algorithm-driven Development application by Team 1}
\label{fig:examples_lvls}
\end{figure}

Team 1 consisted of seven developers, including both junior and senior profiles (three with 1–3 years of experience, two with 8–10 years, and two with over 10 years). They were responsible for both the design and end-to-end implementation of the ADD methodology, acting as its initiators and main practitioners throughout the project. The application developed is a critical, cloud-native component of the \textbf{3D}Experience\textregistered\ platform \citep{dassault3dexperience2024}, used by sales teams to engage with clients, deploy collaborative environments, manage product portfolios, generate custom quotations, and orchestrate cloud provisioning upon payment.
As the originating team behind ADD, they applied the approach consistently over a four-year period. The project focused on the development of a business-critical application, tightly integrated with approximately ten other core services including billing, finance, and infrastructure, while maintaining continuous availability within a cloud environment. This extended case study offers a comprehensive perspective on the potential of ADD when applied end-to-end within a complex, production-grade ecosystem.
To illustrate how ADD supports software engineering across varying levels of complexity, three representative examples from this team are introduced (Figure~\ref{fig:examples_lvls}). Each example reflects a different level of abstraction and technical difficulty. The first example focuses on a basic technical feature, serving to demonstrate the operational steps of ADD when applied to a minimal, well-scoped functionality. The second example addresses a functionality of medium complexity and is intended to show how ADD supports structured planning across multiple sprints while managing inter-team dependencies. The third example describes a complex migration scenario involving numerous distributed services and high delivery constraints, highlighting ADD’s capacity to support robustness, traceability, and fault tolerance in large-scale transformations.

\FloatBarrier

These examples are followed by an analysis of the team’s overall results, including quality metrics, test coverage, and delivery performance. Taken together, they offer insight into the long-term effectiveness of ADD when applied continuously in a production-grade environment.

\subsubsection{Illustrative Example 1: Applying ADD to a Simple Technical Function}
\label{sec:example1}

To illustrate the operationalization of ADD at a low level of abstraction, a basic functionality was selected: enabling users to upload a company logo via a user interface. This feature required the development of a dedicated REST API responsible for validating the request, processing the upload, and ensuring compliance with specified constraints. Despite its technical simplicity, the example effectively demonstrates how ADD supports systematic test derivation from algorithmic design.

The client expressed a set of requirements covering expected behaviors and constraints:

\begin{enumerate}
    \item The user should be able to upload a logo to the company if no logo already exists.
    \item If the logo already exists, then return an error "The logo for this company already exists".
    \item The logo should not exceed 2Mo in size.
    \item The logo must be uploaded in PNG or JPEG format.
\end{enumerate}

The initial step consists in formalizing the client requirements into a structured flowchart, as shown in Figure~\ref{fig:logo}:

\begin{figure}[h]
	\centering
	\includegraphics[width=1\textwidth]{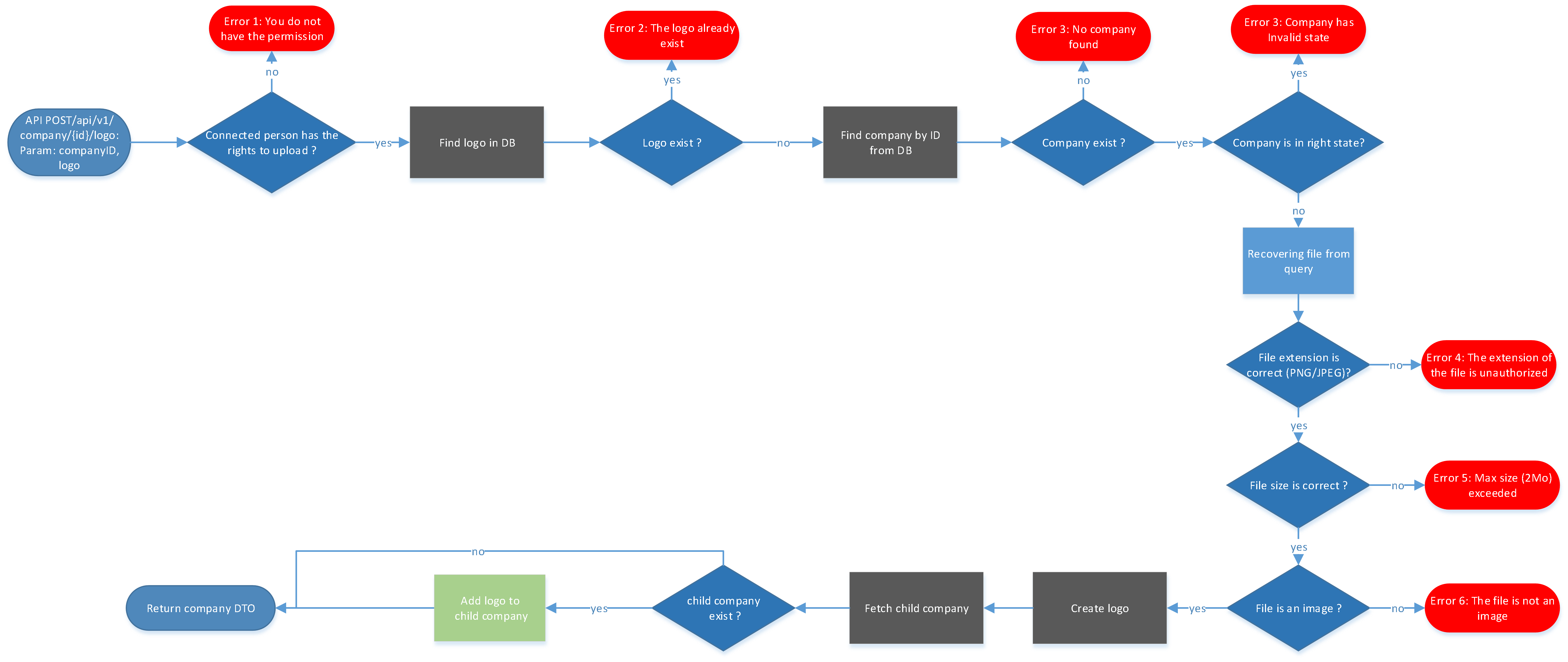}
    \caption{Upload logo flowchart}
    \label{fig:logo}
\end{figure}
\FloatBarrier

During the algorithm design phase, additional edge cases not specified in the initial requirements were identified. For example, the need to validate whether the user has the necessary permissions to modify the logo of the target company emerged as a critical security consideration. This validation was integrated directly into the algorithm, ensuring that such boundary conditions were addressed proactively. This exemplifies how ADD facilitates the discovery and inclusion of overlooked scenarios at an early stage, contributing to system robustness.

Subsequently, the algorithm is analyzed to extract all behavioral paths, each representing a unique expected client scenario. The resulting algorithm contains nine distinct paths through the system. Each path corresponds to a specific behavior and requires a dedicated acceptance test to ensure complete coverage, as shown in Figure~\ref{fig:testsalgo}:

\begin{figure}[h]
	\centering
	\includegraphics[width=1\textwidth]{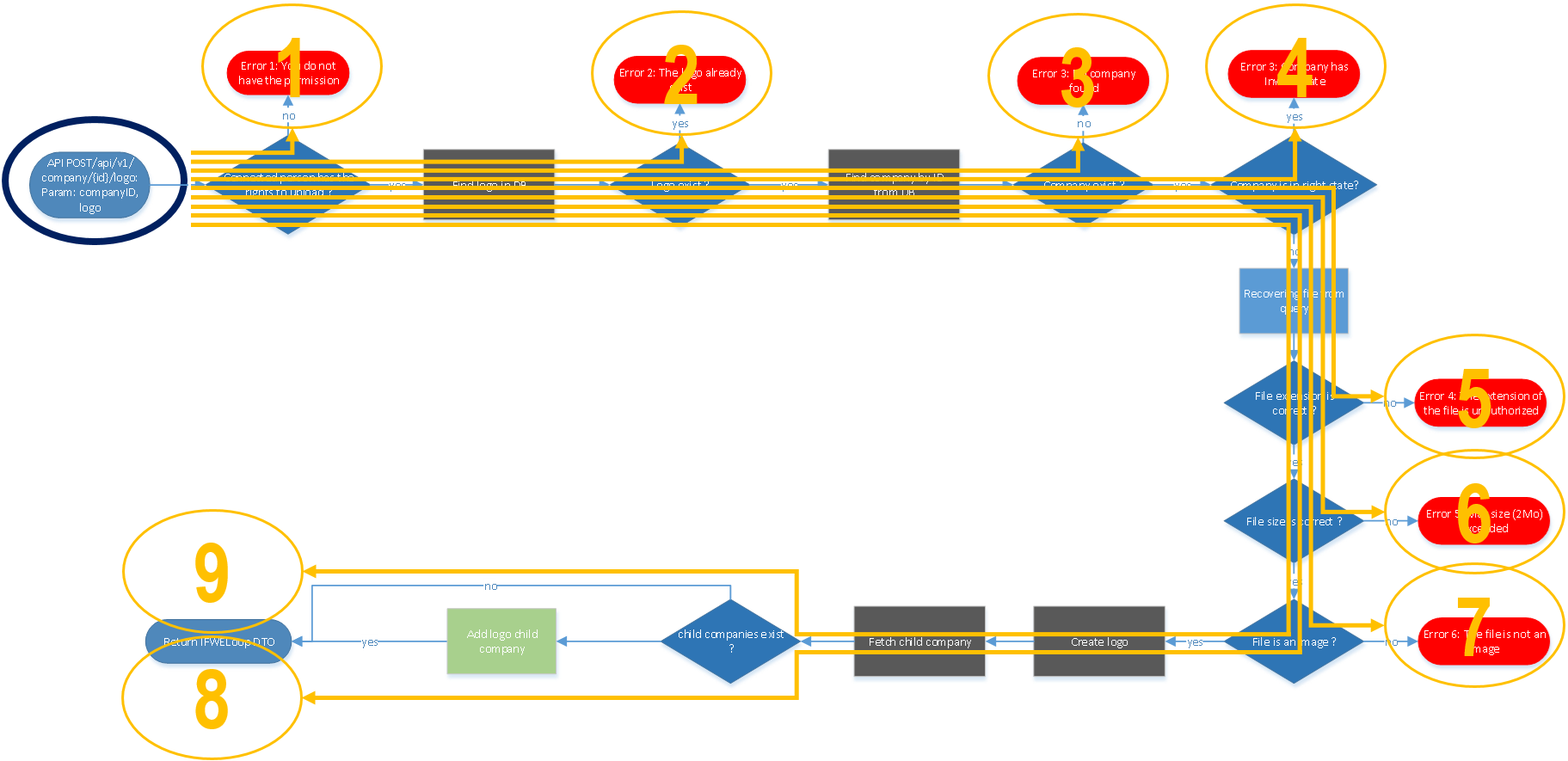}
    \caption{Complete Path Analysis for Acceptance Testing}
    \label{fig:testsalgo}
\end{figure}
\FloatBarrier

For readability, the nine execution paths derived from the flowchart are summarized in Table~\ref{tab:logo_paths}.

\begin{table}[h!]
\centering
\caption{Execution paths derived from the logo-upload flowchart}
\label{tab:logo_paths}
\renewcommand{\arraystretch}{1.2}
\begin{tabularx}{\textwidth}{cXX}
\toprule
\textbf{Path} & \textbf{Condition / branch} & \textbf{Expected behavior} \\
\midrule
P1 & Connected user does not have upload permission & The request is rejected with a permission error. \\

P2 & User has permission, but a logo already exists for the company & The upload is rejected to prevent replacing an existing logo. \\

P3 & No company is found for the provided company identifier & The request stops and returns a company-not-found error. \\

P4 & The company is found but is in an invalid state for logo upload & The request is rejected with an invalid-state error. \\

P5 & The uploaded file has an unauthorized extension & The request is rejected with a file-extension error. \\

P6 & The uploaded file exceeds the maximum allowed size & The request is rejected with a file-size error. \\

P7 & The uploaded file is not recognized as an image & The request is rejected with an invalid-image error. \\

P8 & All validations succeed and no child company is associated & The logo is created and the updated company information is returned. \\

P9 & All validations succeed and a child company is associated & The logo is created, associated with the child company, and the updated company information is returned. \\
\bottomrule
\end{tabularx}
\end{table}

Finally, acceptance tests are designed iteratively based on the identified paths and supported by the diagram’s color-coding convention. Taking the example of path (2) from the flow diagram, the test resembles the structure shown in Figure~\ref{fig:testexample}:

\begin{figure}[h]
	\centering
	\includegraphics[width=1\textwidth]{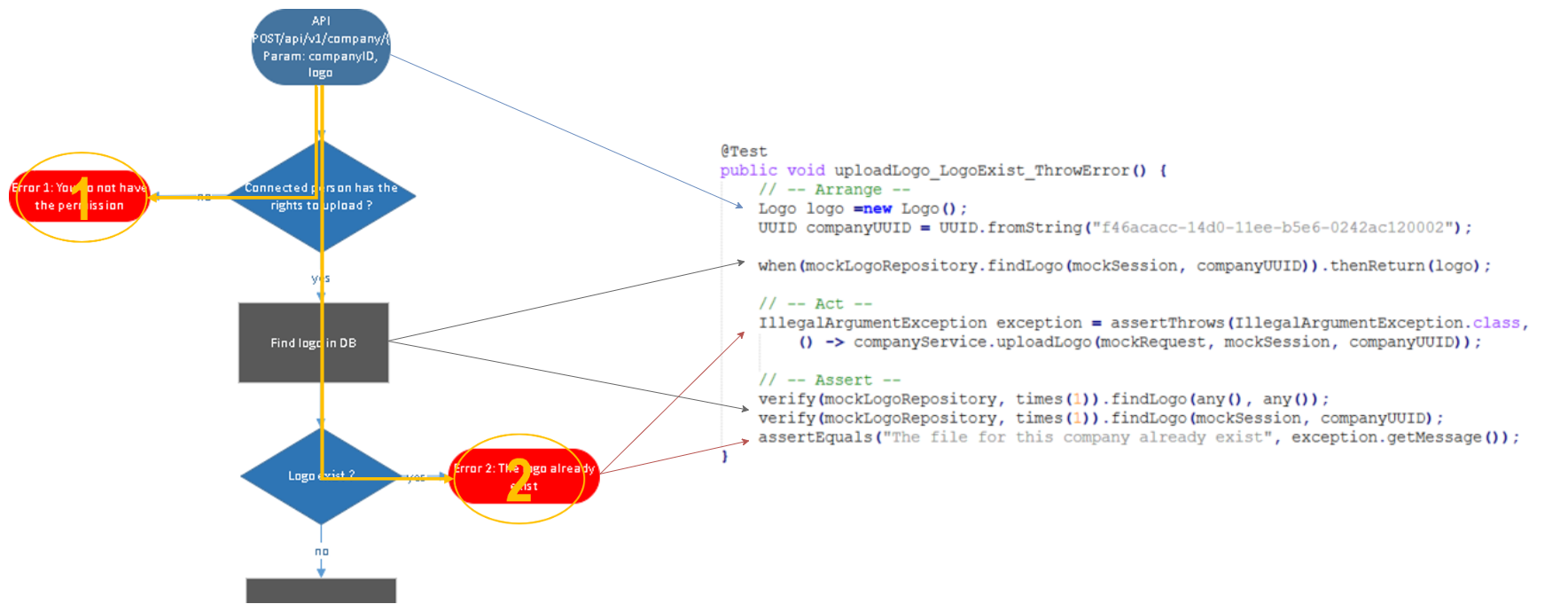}
    \caption{Java test of path (2) from the upload logo diagram}
    \label{fig:testexample}
\end{figure}
\FloatBarrier

By following ADD’s structured approach, both assertions and verifications can be directly extracted from the flowchart. For instance, if the company already has a logo (Error 2), the system must throw an \texttt{IllegalArgumentException} with a precise message. This assertion is explicitly dictated by the flowchart:

\begin{lstlisting}
IllegalArgumentException exception = assertThrows(
    IllegalArgumentException.class, 
    () -> companyService.uploadLogo(mockRequest, mockSession, companyUUID)
);

assertEquals("The file for this company already exists",
             exception.getMessage());
\end{lstlisting}

Beyond assertions, the flowchart also clarifies which interactions must be verified. The "Find logo in DB" step in the diagram represents a call to an external dependency, \texttt{mockLogoRepository.findLogo(...)}. The diagram’s color coding plays a critical role: Table~\ref{tab:color_coding_implication} extends the earlier example and builds upon Table~\ref{tab:color_coding} by summarizing both the semantic meaning and test design implications of each color used in the flowchart.

\begin{table}[h]
\centering
\caption{Flowchart Color-Coding: Semantic Meaning and Testing Implications}
\label{tab:color_coding_implication}
\renewcommand{\arraystretch}{1.8}
\begin{tabularx}{\textwidth}{
  >{\centering\arraybackslash}m{2.5cm} 
  >{\raggedright\arraybackslash}m{6.5cm} 
  >{\raggedright\arraybackslash}m{6.5cm}}
\toprule
\textbf{Color} & \textbf{Meaning in Flowchart} & \textbf{Implication for Test Design} \\
\midrule
\cellcolor{blue!30}\textbf{Blue} &
Indicates that the instruction or sub-method is implemented within the same module. &
Highlights sequential execution, ensuring correct method calls are tested. \\
\cellcolor{darkgraycustom}\color{white}\textbf{Gray} &
Indicates that the instruction or sub-method belongs to another module that communicates with an external service. &
Signals external system interactions, requiring mocks and verification. \\
\cellcolor{brightred}\color{white}\textbf{Red} &
Indicates error returns or exception-handling branches. &
Represents failure scenarios, leading to exception assertions such as \texttt{assertThrows}. \\
\bottomrule
\end{tabularx}
\end{table}

Following this logic, ADD indicates that the database interaction must be verified to confirm that the repository was queried before the exception was thrown:

\begin{lstlisting}
verify(mockLogoRepository, times(1)).findLogo(any(), any()); 
verify(mockLogoRepository, times(1)).findLogo(mockSession, companyUUID);
\end{lstlisting}

Together, these assertions and verifications illustrate how the test structure follows directly from the selected execution path. The error branch determines the expected exception and message, while the gray dependency node indicates the repository interaction that must be verified. This example demonstrates how ADD enables traceability from requirements to execution paths and tests, even in simple features, while encouraging the proactive identification of edge cases during algorithm design.

\subsubsection{Illustrative Example 2: Applying ADD to a Medium Complexity Function}
\label{sec:example2}

Following the pedagogical example provided in the previous section, this second case study aims to demonstrate the applicability of ADD within an agile development process, focusing on the methodology’s ability to support iterative delivery, cross-developer collaboration, and requirement traceability in an industrial context.

In the context of large-scale enterprise software platforms, automating configuration and deployment tasks has become a key concern for improving operational efficiency and minimizing manual errors. A real-world scenario encountered during the deployment process of the \textbf{3D}Experience\textregistered\ platform \citep{dassault3dexperience2024} illustrates this need. When a new client (tenant) is onboarded, several platform services must be initialized with specific default content and configurations. Historically, this initialization process required considerable manual effort and coordination across multiple teams.

To address this challenge, the business required an automated mechanism to streamline and standardize the initialization of newly provisioned tenants. This led to the design and integration of a dedicated API responsible for initializing the content of various core services during deployment. The API encapsulates the initialization logic, ensures consistency across tenant environments, and significantly reduces setup time.

The development of the API was structured into three one-month sprints to allow for progressive delivery of the required features. The developers involved in this implementation had between one and three years of professional experience. For each sprint, the algorithmic modeling required for a given set of use cases was deliberately limited to approximately half a working day per developer. This limited modeling effort provided a system-level understanding that guided implementation and testing, reducing rework and supporting more accurate estimation. In practice, this initial investment was consistently recovered over the course of the sprint, as the structured flowchart clarified edge cases and anticipated integration points early in the cycle. The first sprint was handled by a single developer, while another developer took over for the remaining two sprints. This rotation was made possible without disrupting the development process, thanks to the use of ADD. This approach facilitated a transparent and pragmatic decomposition of the overall functionality into manageable components aligned with sprint objectives, while also ensuring continuity and shared understanding across developers.

Furthermore, ADD supported the seamless handover between developers by providing structured algorithmic descriptions that served both as implementation guides and as lightweight, yet effective, documentation. The methodology also enabled the systematic identification and validation of use cases through these algorithmic artifacts, which served as a shared reference that facilitated interaction with the client whenever clarification or refinement was required. As a result, the methodology contributed to maintaining development continuity, preserving design intent, and meeting quality expectations and delivery timelines, despite personnel rotation during the project.

\begin{figure}[h]
	\centering
	\includegraphics[width=1\textwidth]{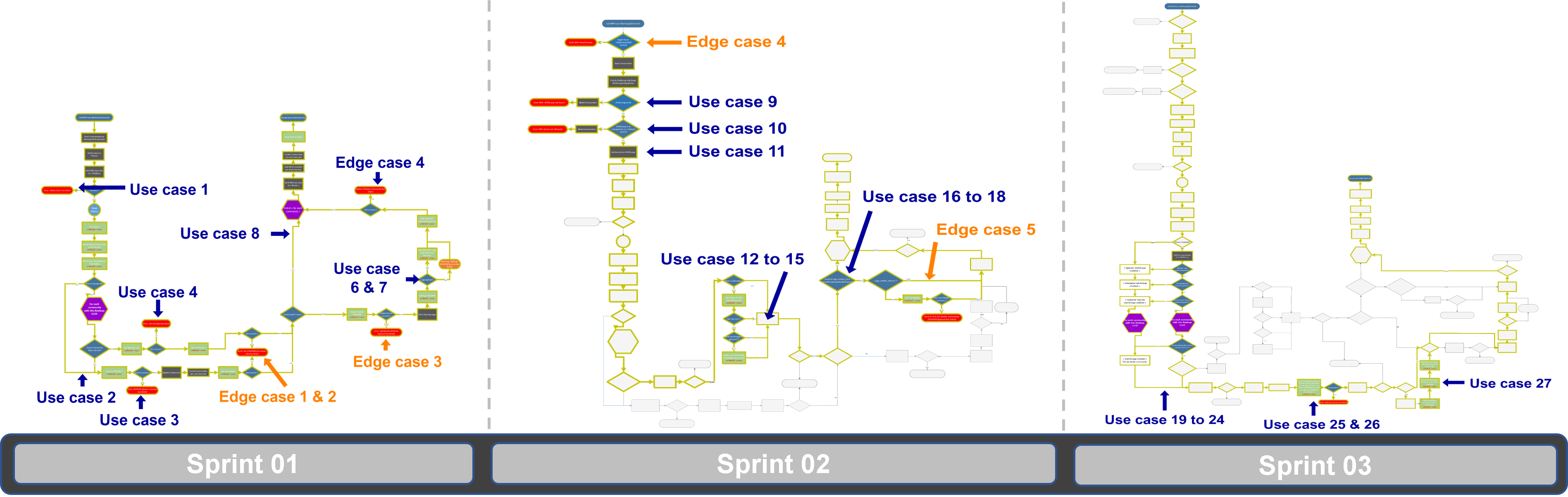}
     \caption{Evolution of the Initialization Algorithm Across Three Development Sprints}
     \label{fig:initialize_overview}
\end{figure}

Figure~\ref{fig:initialize_overview} presents a high-level overview of the algorithm’s evolution across three development sprints. While the content of each block is intentionally blurred to preserve confidentiality, the visual representation emphasizes the incremental nature of ADD. In each sprint, new use cases and edge cases are introduced; newly added branches are shown in color, while pre-existing elements from earlier sprints are grayed out to reflect prior work. Highlighted paths represent newly generated scenarios, which may traverse both new and existing blocks. This layered depiction illustrates how ADD facilitates continuous expansion without disrupting previously validated logic, enabling structured growth and progressive refinement of system behavior.

\FloatBarrier

The timeline of the three sprints illustrates the structured and incremental evolution of the API’s core algorithm, guided by the ADD methodology. In the first sprint, eight primary use cases were implemented, and three edge cases were identified through early client feedback. The second sprint extended functionality with ten additional use cases and revealed two more edge cases requiring clarification. The final sprint completed the intended feature set by incorporating nine remaining use cases, ensuring full alignment with identified edge conditions.

This case study demonstrates how the structured process introduced by ADD supports agile software development. By formalizing requirements into executable flowcharts, integrating early validation through test case extraction, and fostering shared understanding among stakeholders, ADD provides a systematic framework that ensures traceability, facilitates iterative delivery, and promotes technical continuity across evolving team compositions.

Unlike the previous case, which involved short iterative cycles and minimal upfront effort, the next example addresses a high-stakes migration scenario requiring greater investment in modeling and validation. This contrast illustrates the adaptability of ADD across varying delivery constraints and risk profiles.

\subsubsection{Illustrative Example 3: Applying ADD to a complex Migration}
\label{sec:example3}

This example illustrates the feasibility of the ADD methodology when applied to a complex project, such as the migration of clients to a cloud platform. It is important to note that this migration project represents just one of many features developed using ADD within the service. Over the course of four years, the ADD methodology was implemented across the entire service, covering numerous functionalities in a fully agile environment. This example serves to demonstrate how the core principles of ADD can be implemented effectively in a mission-critical context. This migration project was chosen because it reflects the challenges of managing numerous dependencies and coordinating between different services. The following section presents the main challenges, planning approach, execution process, and outcomes of this migration project. The broader context and quality metrics discussed in the subsequent sections reflect the full scope of the team's work with ADD, beyond just the migration project.

\subsubsection*{Project Overview and Challenges}

Dassault Systèmes \citep{dassault2024} is a global leader in developing 3D design, Product Lifecycle Management (PLM), and digital experience software. Its cloud platform the "\textbf{3D}Experience\textregistered\ platform" \citep{dassault3dexperience2024} serves thousands of enterprises worldwide, providing solutions that help companies across industries design, produce, and manage complex products and systems efficiently.
Dassault Systèmes faced a very challenging project which involved migrating over 3,000 enterprise platforms from one cloud solution to another. This migration was not only about transferring data; it involved moving entire platforms that hosted thousands of users and various types of data, including documents, videos, text posts, and other media. The complexity was further compounded by the need to orchestrate numerous APIs to interact with the various services running on these platforms. This complex client migration project was one of the critical tasks handled by the development team as part of a larger application initiative. The results discussed in subsequent sections represent the cumulative impact of all such critical tasks, including this migration. Given the critical nature of the migration, no errors were tolerable. The migration process had to be idempotent, meaning that in the event of any interruption, it needed to resume seamlessly from the point of failure without causing inconsistencies \citep{kleppmann2017designing}. This task introduced considerable complexity, as many of these migrations could last for hours, depending on the volume of data. In addition to these complexities, cloud platform migrations pose unique challenges, including managing data consistency across distributed systems, ensuring minimal downtime during transition, maintaining data security during transfer, and orchestrating numerous interdependent services and APIs, each of which requires careful synchronization \citep{zhao2014strategies}. Ensuring data integrity and consistent user experience during this prolonged migration was a top priority.

\subsubsection*{Planning and Implementation with ADD}
This collaboration, supported by the ADD approach, allowed the team to create detailed flowchart diagrams to map out each stage of the migration and ensure that all possible scenarios were considered. During these iterations, the service owners, under the guidance of the two developers, reviewed the relevance of the acceptance tests to ensure their validity and suitability for the migration process.

The ADD modeling phase, including flowchart design and validation of acceptance criteria, was conducted over a four-week period. This duration was tightly integrated into the project's sprint planning and was aligned with the high complexity and criticality of the task. Unlike conventional development workflows where implementation may begin after minimal upfront planning, this preparatory effort significantly reduced the risk of failure. Beginning development earlier would likely have led to costly production issues, increased support workload, and decreased client satisfaction. Within the context of a four-month project, this investment contributed to a production rollout with zero incidents, an outcome of particular value given the stringent delivery expectations. The refinement of the algorithm through collaborative reviews with the Cloud Run manager and the lead architect was instrumental in addressing key concerns such as idempotency, performance, and scalability. Implementation commenced only after full validation of the flowchart and underlying algorithm.

Once the flowchart and algorithm were validated, the ADD test-selection heuristic was applied to select 91 acceptance tests covering the relevant execution paths, decision outcomes, and error scenarios identified in the migration model. The subsequent implementation resulted in 4,120 lines of functional code and 9,802 lines of test code. These acceptance tests were directly derived from the flowchart diagrams, demonstrating the strength of ADD in helping design tests that are aligned with both the client’s needs and the system’s expected behaviors.

This four-week modeling phase helped establish a shared understanding of the migration process among the involved teams, supported the identification of edge cases, and guided the creation of a robust set of acceptance tests derived from the flowchart diagrams. Once the development was completed, the QA phase began. During this phase, test data similar to production data was created, and the QA team used the algorithm as a basis, with assistance from the developers, to ensure no edge cases were missed. On large-scale and high-impact projects, collaboration among different technical teams is crucial. The ADD methodology provided visual support that facilitated exchanges, collaboration, and convergence towards a resilient solution.

\begin{figure}[h]
	\centering
	\includegraphics[width=0.9\textwidth]{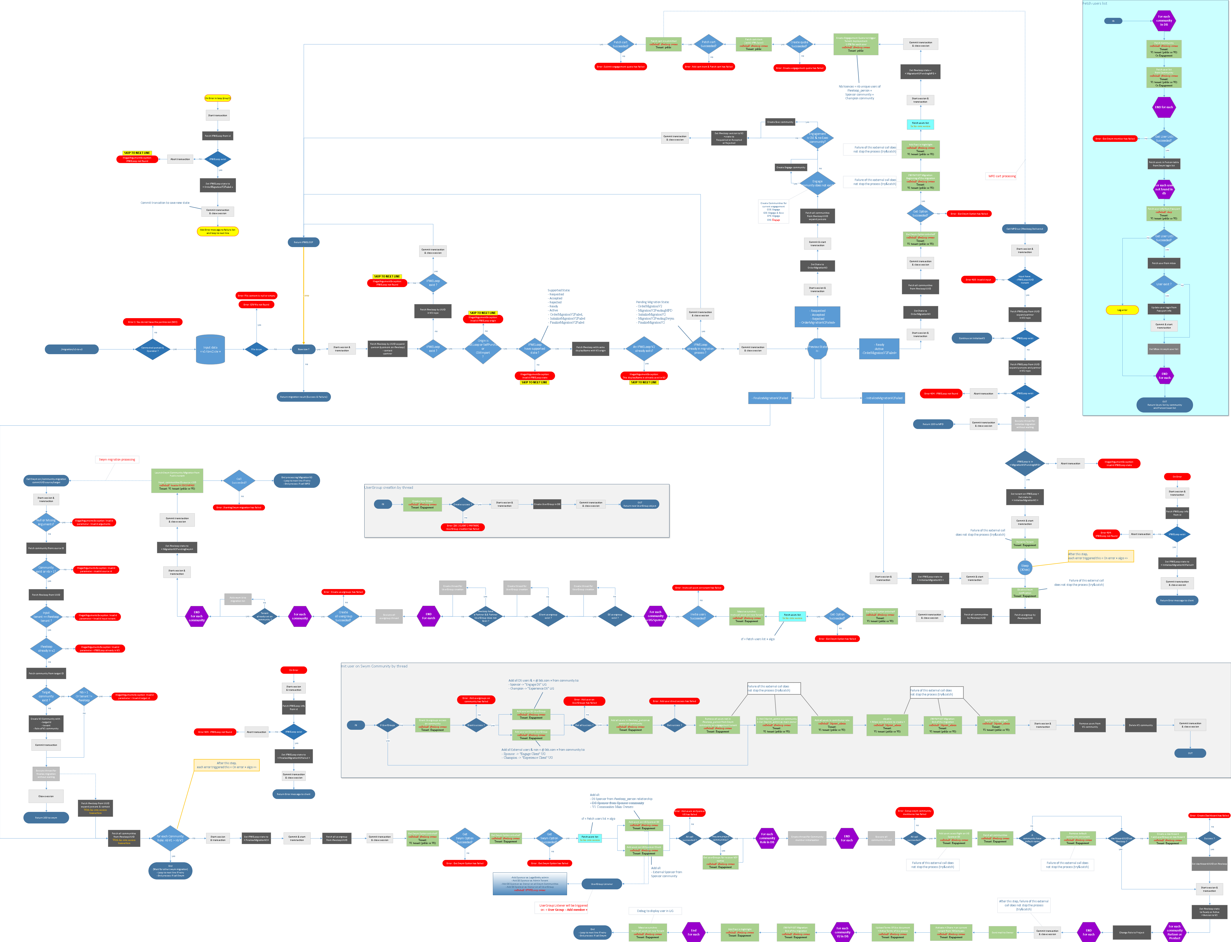}
     \caption{Migration algorithm}
     \label{fig:migration_algo}
\end{figure}

The following diagram (Figure~\ref{fig:migration_algo}) provides an overview of the algorithm. The text is intentionally blurred to preserve confidentiality; however, the model can still be characterized quantitatively. The flowchart had a cyclomatic complexity of 36 (see Section~\ref{sec:QuantitativeResults} for the definition and calculation procedure), indicating at least 36 linearly independent execution paths to be covered, and integrated 49 external APIs/services. Exhaustive enumeration of all possible combinations would have produced a substantially larger test set.

\FloatBarrier

\subsubsection*{Execution and Results}
An additional layer of complexity came from the user notification process. Given the scale of the migration, with platforms hosting thousands of users, it was essential to automate communication. Automated scripts were implemented to send pre-migration notifications to users, informing them of the upcoming migration and providing an estimated completion time based on the volume of data to be migrated. Users were kept in the loop at all stages, including receiving a post-migration notification confirming that the migration was successful and inviting them to report any issues they might notice. Automated user notifications were a key element in maintaining transparency throughout the migration. These notifications helped reduce uncertainty among users by providing clear information about the migration timeline and expected impacts, which significantly minimized disruptions and improved the overall user experience.

This extensive testing phase ensured that the results were highly successful: over 3,000 enterprise platforms were migrated without a single issue in production. The use of ADD played a critical role in this success, as also reflected in the feedback provided by QA, operations, and sales stakeholders during and after the migration. Once the migration was underway, only one person was needed on support to monitor the progress by checking the generated reports, ensuring that everything proceeded as planned.

Although the core development team was directly involved in the application of ADD, several aspects of the approach were independently validated by external stakeholders throughout the project. The QA team confirmed that the algorithmic documentation significantly facilitated the identification of edge cases and accelerated the preparation of test data, including scenarios involving corrupted or incomplete records. From an operational standpoint, the migration process received positive feedback from department leadership, particularly the QA lead and the cloud operations team, who emphasized the absence of incidents and the smooth coordination across services. In addition, the sales leadership team responsible for clients accounts management expressed strong satisfaction following the successful migration of thousands of enterprise platforms without disruption, highlighting the positive impact on client trust and post-migration engagement. These validations reinforce the practical relevance and cross-functional value of ADD beyond the development team itself.

\subsubsection*{Value of Proactive Planning}
This reduction in support requirements during production was a direct result of the meticulous preparation enabled by ADD. Without the detailed flowcharts to visualize the entire process and ensure that every edge case was covered, it is likely that more significant resources would have been required for support, especially for handling unforeseen issues.

The careful planning and proactive testing made possible by ADD ensured that the entire migration process, from planning and designing the algorithm to implementing and testing, was completed with zero errors and minimal overhead during production. This upfront investment in algorithm design and test preparation led to significant savings and improved reliability during the production phase, illustrating the practical benefits of ADD.

Across the three scenarios, the ADD methodology demonstrated flexibility in effort allocation, ranging from half-day modeling in agile sprints to full-scale preparatory design in mission-critical settings. In all cases, the approach helped surface edge cases early, align teams, and reduce downstream rework.

\subsubsection{Team 1: Overall Team Performance and Improvement Metrics}
In Team 1, the introduction of ADD significantly improved the clarity of requirements and the overall development process. Early use of flowcharts prompted essential discussions with the specification team, particularly to resolve ambiguities and define edge cases before implementation began. This proactive refinement ensured that development was closely aligned with client expectations.

The diagrams also served as accessible, evolving documentation. Developers unfamiliar with a feature could rely on the diagrams to quickly grasp its structure and adapt them as specifications evolved, minimizing the need for separate documentation efforts.

As a result, the team achieved a steady delivery rhythm consistent with agile principles~\citep{beck2001agile}, along with more accurate sprint estimates and a reduced defect rate. Compared to prior TDD-based workflows, ADD supported better planning, stronger alignment between tests and features, and a clearer development process.

The performance metrics were drawn from four years of production data collected through Dassault Systèmes' internal tools, including defect tracking, test cycles per sprint, and productivity data based on Git commits. Test coverage was assessed via CI/CD pipelines using JUnit (backend) and Page Object frameworks (frontend and integration).

\subsubsection{Defects per lines of code}

The project consisted of 22,444 lines of code and a total of 110 defects identified during in-house testing, with only 2 defects reported after release extracted from the organization’s internal Application Lifecycle Management (ALM) system. This corresponds to a defect density of approximately 4.90 defects per KLOC during the QA phase and 0.089 defects per KLOC post-release, measured across the entire lifecycle of the project. Because the post-release density is based on only two observed defects, this value should be interpreted as a descriptive indicator rather than as a statistically robust estimate.

To contextualize these values, Shah et al. \citep{shah2012overview} conducted a scoping study that aggregates defect density data from 57 primary studies published between 1992 and 2010. Within the subset of projects developed in Java, the study reports a mean defect density of 5.9 defects/KLOC, although it does not consistently specify whether this value reflects the testing or post-release phase. Given this ambiguity, comparisons with phase-specific measurements must be interpreted cautiously, as QA-phase and post-release defect densities quantify different classes of risk and are not directly interchangeable.

Relative to this academic reference point, the QA-phase defect density observed in the present project (4.90 defects/KLOC) is lower, and the post-release defect density (0.089 defects/KLOC) is lower. These values should be interpreted in light of differences in project complexity, application domain, development process, team size, and tooling, as well as the age of the benchmark, whose data collection period predates the widespread adoption of contemporary CI/CD, automated testing, and cloud-native development practices.

Although the projects included in Shah et al.’s survey vary in complexity and application domain, the comparison remains meaningful given the shared use of Java and the inclusion of business-oriented and web-based applications. The studied project, a cloud-based application written in Java, aligns with several systems represented in the reference dataset, although differences in development processes, team size, and tooling may still influence outcomes.

\begin{figure}[h]
	\centering
	\includegraphics[width=0.5\textwidth]{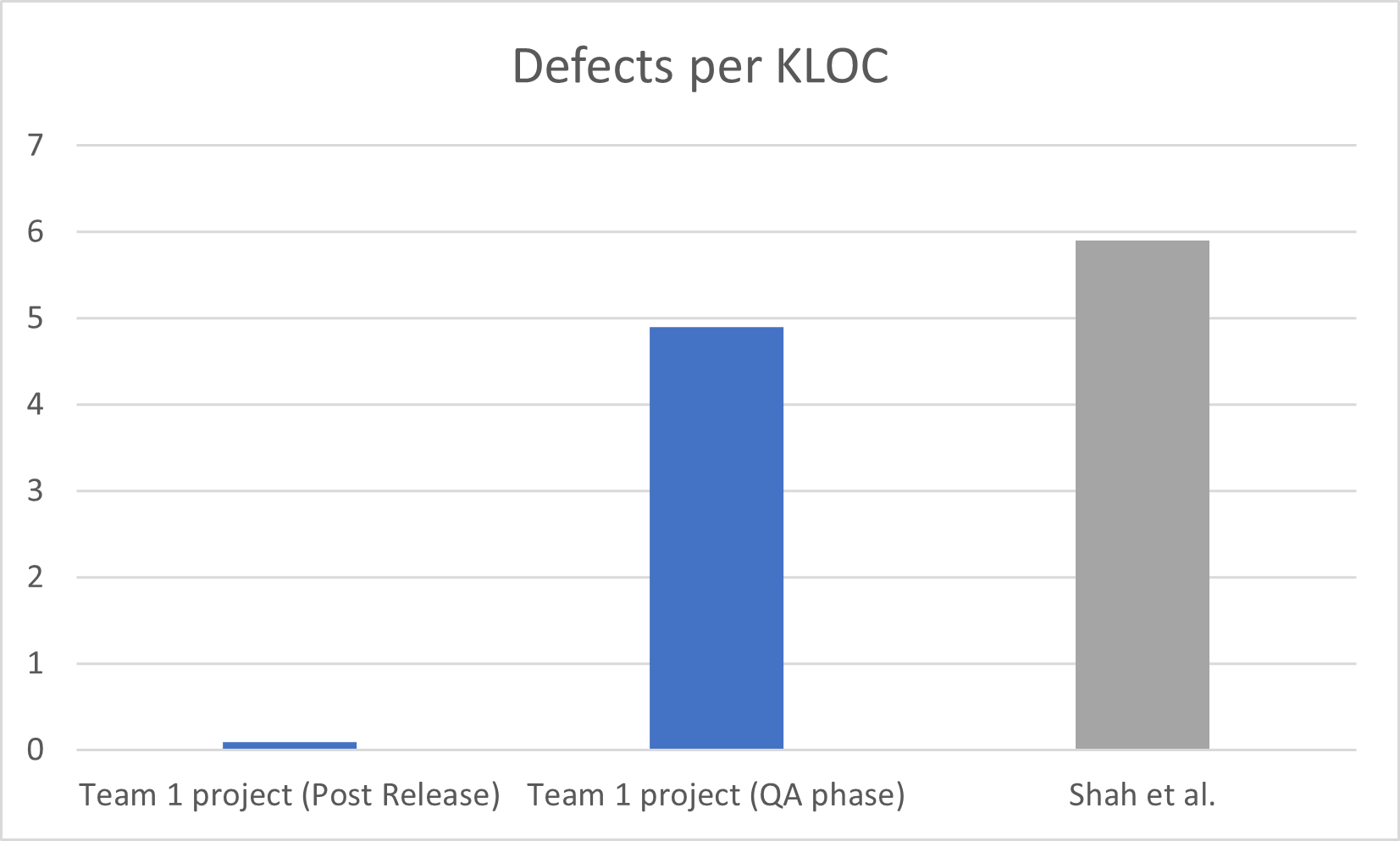}
    \caption{Defects per 1000 lines of code compared to Shah et al. study}
    \label{fig:bug_perKloc}
\end{figure}

Figure~\ref{fig:bug_perKloc} illustrates the observed defect densities in comparison with the range of values reported for Java-based systems in the scoping study.

\FloatBarrier

An analysis of the two recorded defects revealed that both originated from scenarios not represented in the initial algorithmic diagrams. In one case, a rare edge condition was overlooked during the modeling phase, preventing its detection during testing. In the other, an environment-specific configuration issue emerged, which had not been reproducible under the controlled testing conditions. These occurrences illustrate the dependency of ADD’s effectiveness on the completeness of the diagrams and the coverage of the test environment.

\subsubsection{Coverage}
The reported code coverage values were automatically collected from the CI/CD pipeline used by the development team. 
Coverage metrics were computed continuously during the build and testing process, ensuring that the values reported accurately reflect the actual and up-to-date state of the project throughout its lifecycle. As these metrics were automatically extracted from the CI/CD pipeline at every build, the reported coverage reflects a longitudinal measurement across the entire four-year period, rather than a single snapshot. 
Using this approach, the team consistently maintained a coverage level of 95\% over a four-year period. 
To provide external context for these values, the study by Hilton et al.~\citep{hilton2018large} analyzed 47 software projects across various sectors and reported an average code coverage between 74\% and 76\%. 
It is important to note that these values represent the state of the projects at the time of measurement and do not necessarily reflect industry satisfaction or the absence of further testing efforts. 
In many cases, teams may have aimed for higher coverage but faced organizational or technical constraints. 
In contrast, the sustained 95\% coverage achieved in this project reflects a deliberate strategy to enforce comprehensive testing, including unit, integration, and acceptance tests, thus providing a robust safety net against regressions and newly introduced defects.

\begin{figure}[h]
	\centering
	\includegraphics[width=0.5\textwidth]{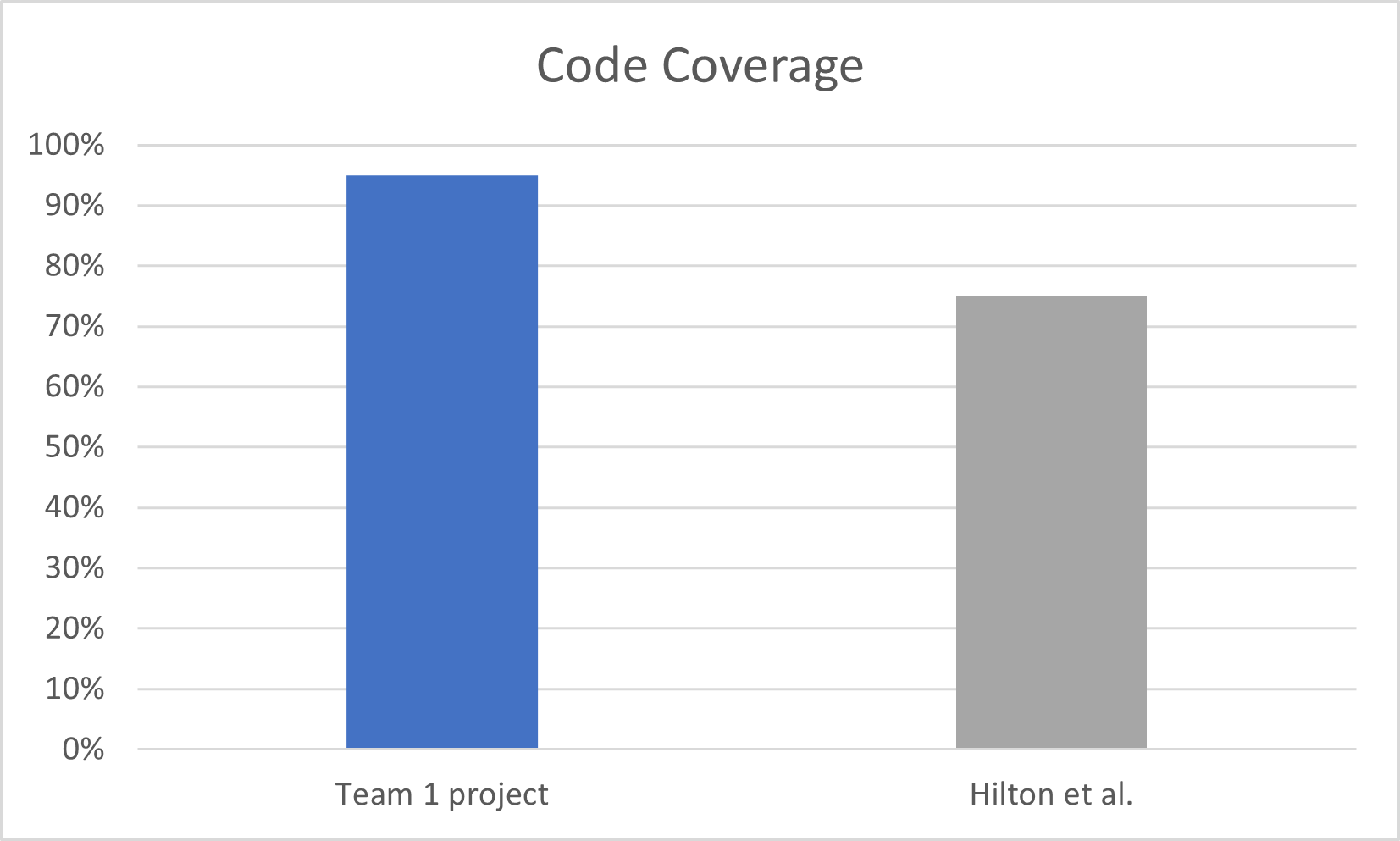}
    \caption{Comparison of Team~1 code coverage with industry averages}
    \label{fig:coverage}
\end{figure}

Figure~\ref{fig:coverage} provides a visual comparison between the code coverage consistently maintained by Team~1 over four years and the average coverage values reported by Hilton et al.~\citep{hilton2018large}. 
This representation highlights the gap between the deliberate testing strategy adopted in this project and the broader industry benchmarks.

\FloatBarrier

\subsubsection{Delivery Cadence with ADD}

\begin{figure}[h]
    \centering
    \includegraphics[width=\textwidth]{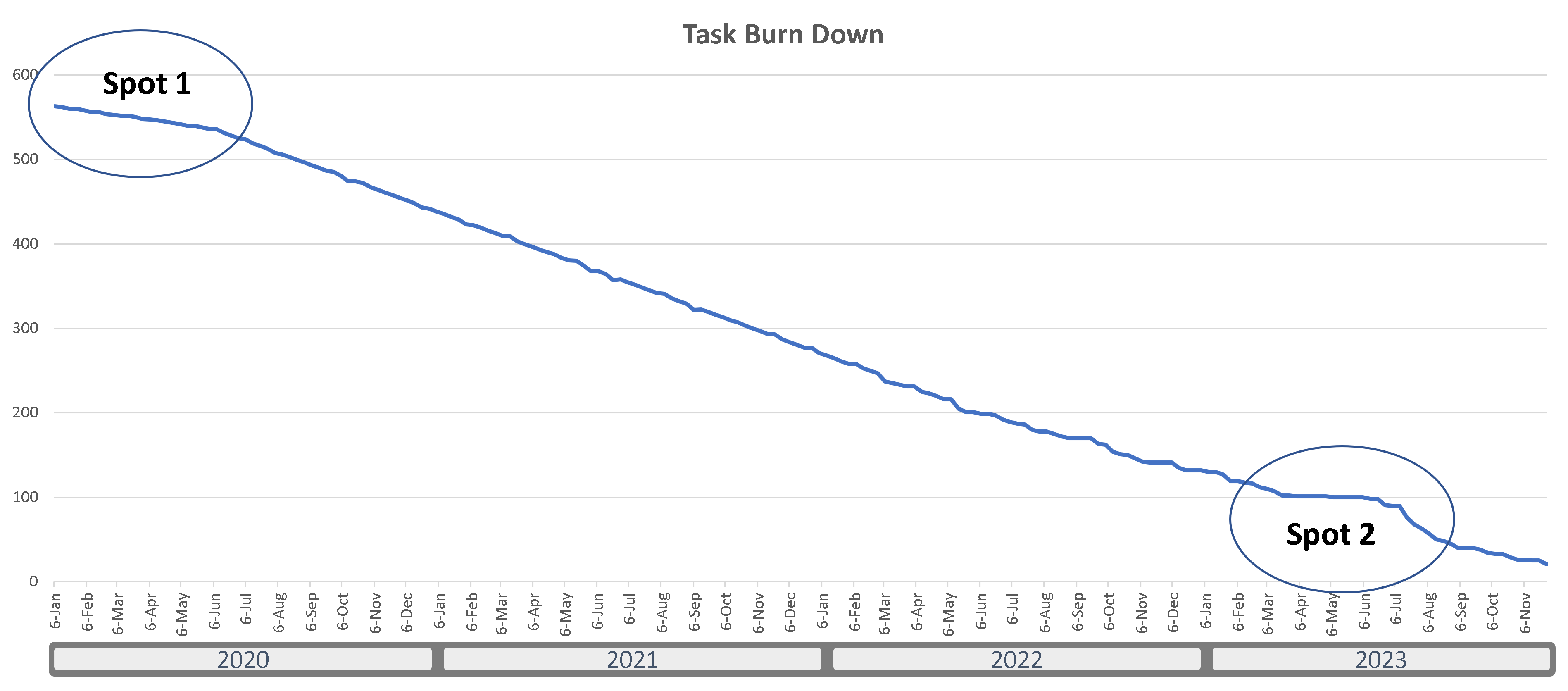}
    \caption{Four-year burn down chart for Team~1, extracted from the ENOVIA Project Planning application.}
    \label{fig:cadence_team1_enovia_planner}
\end{figure}

To evaluate the long-term stability of delivery pace in the context of the ADD approach, task completion data were obtained from the ENOVIA \textit{Project Planner} \citep{DassaultENOVIAProjectPlanner} application. This tool automatically generates burn down charts by aggregating data from the project backlog, where each task corresponds to a user story or development work item recorded in the system. Data points represent the total number of open tasks at regular intervals, providing a direct view of delivery cadence over time.
To ensure comparability over the entire observation period, larger tasks were systematically decomposed so that all backlog items maintained a similar level of granularity. As illustrated in Figure~\ref{fig:cadence_team1_enovia_planner}, the resulting burn down chart for Team~1 covers a four-year period.

\FloatBarrier
Two distinct inflection periods can be identified in the delivery cadence.
The first, an initial plateau (Spot 1), occurs at the very beginning of the project and is characterised by a relatively shallow slope, indicating a slower visible delivery rate. During this stage, substantial effort was devoted to establishing the development environment, including repository creation, framework integration, testing infrastructure setup, and key technical decisions. Although these activities did not immediately reduce the backlog count, they were essential to enabling sustained development in subsequent phases.

The second inflection, a mid-project plateau (Spot 2), corresponds to a deliberate slowdown initiated by the business organisation to prioritise partner onboarding on the platform before resuming feature delivery. During this interval, the development team continued implementing planned features in the background. The sharp decline in the backlog immediately after this plateau reflects the release of this accumulated work once the business constraints were lifted.

Following the second plateau, the team maintained a steady delivery rate, closing the remaining backlog in preparation for key milestones.

Overall, the longitudinal data show that the application of ADD enabled the team to maintain a nearly linear delivery cadence over multiple years. This sustained rhythm made it possible to practice agility at scale, with sprint plans that were both realistic and consistently met. Business and QA teams reported particularly positive feedback on the predictability and reliability of delivery, which supported efficient cross-team coordination and long-term planning. This longitudinal stability in delivery cadence provides empirical evidence that ADD not only supports short-term sprint planning but also sustains agility at scale in complex, multi-year projects.

These results further support the claim that ADD fosters not only short-term quality improvements but also sustainable delivery performance in long-lived industrial projects.

\subsection{Team 2: Application of ADD in a critical environment}
Team~2 was composed of fifteen software developers responsible for a core infrastructure service of the \textbf{3D}Experience\textregistered\ platform. This service manages user authentication, licensing, and role-based access control across multiple interconnected applications, ensuring secure and seamless interoperability. Given its central role, its reliability and maintainability are essential to the platform’s performance. While the platform incorporates multiple layers of redundancy and failover mechanisms to ensure continuous availability, any defect in this service could require rapid remediation to prevent functional disruptions.

The team included a balanced mix of junior (1 to 3 years of experience), intermediate (4 to 9 years), and senior (10 or more years) developers, with experience levels ranging from 1 to 17 years. Prior to the introduction of ADD, the team alternated between a test-last approach, writing tests only when time allowed, and a partial adoption of TDD for selected features. In this study, “test-last development” refers to an approach in which testing is typically performed after implementation and not always systematically.

During an 18-month complete redesign of the service, ADD was progressively introduced with structured training and continuous support from Team~1, who had already implemented ADD successfully in another project. This case provides a unique opportunity to examine the adoption of ADD in a mature, high-stakes industrial setting, and to quantify its impact on both productivity and software quality.

\subsubsection{Training and Methodology Adoption}

As part of the complete redesign of the service, Team~1 was requested to provide methodological support to Team~2. This support included the introduction of ADD and a structured training program designed to enable the team to apply the methodology effectively within their context.

The onboarding sequence consisted of three phases. First, all developers participated in a full-day theoretical session covering the foundations of ADD, its process stages, and its intended benefits compared to existing practices. This was followed by a half-day practical workshop in which participants worked through example cases, producing algorithmic models and associated test plans. Finally, each developer group received one week of hands-on coaching during the implementation of a real use case from their own service. This practical phase was intended to ensure the transfer of skills to actual production work.

Long-term adoption was supported by ongoing oversight from technical leads, who periodically reviewed the produced algorithmic models and corresponding code to verify correct application of ADD principles. This monitoring aimed to prevent gradual abandonment of the methodology, a risk observed in similar process-change initiatives \citep{anastassiu2020resistance}.

The extent and patterns of adoption across experience levels are analyzed in the results section, alongside quantitative measures of delivered features and associated defect rates.

\subsubsection{Quantitative Results}
\label{sec:QuantitativeResults}

To assess the impact of ADD on software quality, all functions delivered during the 18-month project were analyzed. Data was extracted from the organization’s internal ALM system, which integrates task tracking, feature delivery records, and defect reports. For each function, the ALM provided the assigned developer, development methodology used, estimated complexity, and all associated defect reports identified either during internal QA testing or post-release. Only defects explicitly linked to the function’s internal logic were considered in this study.

Each function was assigned a complexity score using estimated cyclomatic complexity. For flowchart-modeled components, complexity was calculated by counting decision nodes (e.g., diamonds). For code-based components, a static scan was performed by counting decision statements such as \texttt{if}, \texttt{else}, and \texttt{while}.

Cyclomatic complexity~\citep{mccabe1976complexity} is defined as:
\begin{equation}
CC = E - N + 2P
\end{equation}
where $CC$ is the cyclomatic complexity, $E$ is the number of edges in the control flow graph, $N$ is the number of nodes, and $P$ is the number of connected components (typically 1 for a single function). For practical analysis of individual functions, this formula is often simplified by directly counting decision points such as \texttt{if}, \texttt{else}, \texttt{while}, and \texttt{for} constructs. The simplified form used in this study is:
\begin{equation}
CC = \text{Number of decision points} + 1
\end{equation}

Functions were categorized into three complexity levels based on commonly used interpretations of McCabe's cyclomatic complexity. The threshold of 10 has historically been used as a practical upper bound for acceptable module complexity in structured testing \citep{watson1996structured}. Cyclomatic complexity also remains widely used in recent empirical studies as a code characteristic related to understandability and maintainability \citep{lavazza2023empirical}. In this study, we therefore used three broad categories to compare defect rates across increasing levels of control-flow complexity:

\begin{itemize}
  \item \textbf{Low}: $1 \leq CC \leq 10$
  \item \textbf{Medium}: $11 \leq CC \leq 20$
  \item \textbf{High}: $CC > 20$
\end{itemize}

\begin{figure}[h]
	\centering
	\includegraphics[width=1\textwidth]{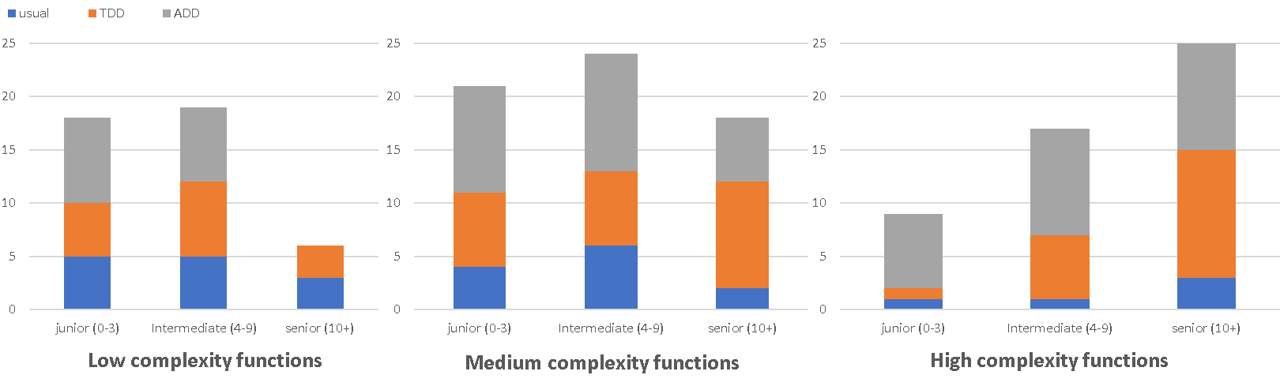}
     \caption{Distribution of delivered features by development method and feature complexity across developer experience levels.}
     \label{fig:team2_functionComplexityRepartition}
\end{figure}

The bug rate was then computed for each complexity level and development approach (test-last, TDD, ADD). In total, 157 API functions were analyzed, comprising 51 high-complexity functions, 63 medium-complexity functions, and 43 low-complexity functions. Each function was tagged with the development methodology used, enabling a detailed comparison of defect rates relative to both functional complexity and coding practices. The distribution of these functions by method and developer is presented in Table~\ref{tab:team2}, the corresponding breakdown by experience level is illustrated in Figure~\ref{fig:team2_functionComplexityRepartition}, and the average number of bugs per function is summarized in Table~\ref{tab:bugs_comparison}.

\FloatBarrier
\begin{table*}[t]
\scriptsize
\centering
\caption{Number of delivered features and associated bugs by developer, method, and complexity (Team 2)}
\label{tab:team2}
\rowcolors{2}{gray!8}{white}
\begin{tabularx}{\textwidth}{c c *{9}{>{\centering\arraybackslash}X}}
\toprule
\rowcolor{white}
\textbf{Dev ID} & \makecell{\textbf{Exp.}\\\textbf{(yrs)}} & 
\makecell{\textbf{Test-last}\\\textbf{Low}} & \makecell{\textbf{Test-last}\\\textbf{Med}} & \makecell{\textbf{Test-last}\\\textbf{High}} & 
\makecell{\textbf{TDD}\\\textbf{Low}} & \makecell{\textbf{TDD}\\\textbf{Med}} & \makecell{\textbf{TDD}\\\textbf{High}} & 
\makecell{\textbf{ADD}\\\textbf{Low}} & \makecell{\textbf{ADD}\\\textbf{Med}} & \makecell{\textbf{ADD}\\\textbf{High}} \\
\midrule
1  & 10 & 0       & 1 (2)  & 1 (9)  & 1 (0)  & 2 (3)  & 3 (12) & 0     & 1 (0)  & 2 (1)  \\
2  & 5  & 1 (4)   & 1 (4)  & 0      & 1 (1)  & 2 (2)  & 1 (5)  & 2 (0) & 2 (1)  & 1 (1)  \\
3  & 17 & 0       & 1 (4)  & 1 (7)  & 1 (1)  & 2 (1)  & 3 (10) & 0     & 1 (1)  & 2 (2)  \\
4  & 17 & 1 (3)   & 0      & 1 (8)  & 0      & 2 (2)  & 2 (7)  & 0     & 1 (0)  & 2 (1)  \\
5  & 11 & 1 (4)   & 0      & 0      & 1 (1)  & 1 (2)  & 1 (4)  & 0     & 2 (1)  & 3 (2)  \\
6  & 5  & 1 (3)   & 1 (5)  & 0      & 2 (3)  & 1 (2)  & 1 (5)  & 1 (0) & 2 (1)  & 2 (2)  \\
7  & 2  & 1 (4)   & 1 (4)  & 0      & 1 (2)  & 1 (3)  & 1 (6)  & 2 (0) & 3 (2)  & 2 (2)  \\
8  & 5  & 1 (3)   & 1 (4)  & 0      & 2 (1)  & 1 (2)  & 0      & 1 (0) & 2 (1)  & 2 (2)  \\
9  & 7  & 1 (2)   & 1 (5)  & 1 (10) & 1 (1)  & 1 (2)  & 2 (9)  & 2 (0) & 3 (1)  & 3 (3)  \\
10 & 3  & 1 (3)   & 1 (3)  & 0      & 1 (1)  & 2 (3)  & 0      & 2 (0) & 2 (0)  & 1 (3)  \\
11 & 6  & 1 (3)   & 2 (6)  & 0      & 1 (0)  & 2 (2)  & 2 (9)  & 1 (0) & 2 (2)  & 2 (1)  \\
12 & 17 & 1 (3)   & 0      & 0      & 0      & 3 (2)  & 3 (10) & 0     & 1 (0)  & 1 (0)  \\
13 & 3  & 1 (2)   & 1 (4)  & 0      & 1 (2)  & 1 (2)  & 0      & 2 (0) & 2 (1)  & 1 (2)  \\
14 & 1  & 1 (3)   & 1 (5)  & 1 (10) & 1 (1)  & 1 (3)  & 0      & 1 (0) & 2 (1)  & 1 (1)  \\
15 & 3  & 1 (2)   & 0      & 0      & 1 (0)  & 2 (3)  & 0      & 1 (0) & 1 (0)  & 2 (2)  \\
\midrule
\rowcolor{gray!18}
\textbf{Total} & \textbf{--} & 
\textbf{13 (39)} & \textbf{12 (46)} & \textbf{5 (44)} &
\textbf{15 (14)} & \textbf{24 (34)} & \textbf{19 (77)} &
\textbf{15 (0)} & \textbf{27 (12)} & \textbf{27 (25)} \\
\bottomrule
\end{tabularx}
\rowcolors{2}{white}{white}
\end{table*}
\FloatBarrier

\FloatBarrier

\begin{table}[h!]
\centering
\caption{Average number of bugs per function, with number of functions in parentheses}
\label{tab:bugs_comparison}
\renewcommand{\arraystretch}{1.3}
\begin{tabularx}{\textwidth}{lCCC}
\toprule
\textbf{Complexity} & \textbf{Test-last} & \textbf{TDD} & \textbf{ADD} \\
\midrule
Low    
& 3 bugs/function (13)  & 1 bug/function (15)     & 0 bugs/function (15) \\
Medium 
& 4 bugs/function (12)  & 1.5 bugs/function (24)  & 0.5 bugs/function (27) \\
High   
& 9 bugs/function (5)   & 4 bugs/function (19)    & 1 bug/function (27) \\
\bottomrule
\end{tabularx}
\end{table}
\FloatBarrier

These results should be interpreted as descriptive industrial observations rather than inferential statistical evidence. No statistical significance test was performed, as the data were collected in an operational setting and the functions were not assigned to development approaches through a controlled experimental design. The values reported in Table~\ref{tab:bugs_comparison} therefore indicate observed trends within the studied context. In particular, the consistent decrease in the average number of bugs from test-last development to TDD and ADD across all complexity levels suggests a positive association between ADD adoption and lower observed defect counts, but should not be interpreted as a causal or statistically generalizable effect.
Two additional factors should be considered when interpreting the Team~2 comparison. test-last development and TDD were already part of Team~2's existing practices before ADD was introduced. During the observed period, ADD was then requested from all developers, regardless of experience level and function complexity, in order to collect feedback on the method across different profiles and development situations. However, this introduction was conducted in an operational project setting rather than through a randomized experimental design. In addition, ADD adoption was supported through coaching, review meetings, and monitoring of method application, and this structured support may also have contributed to the observed outcomes. Accordingly, the Team~2 results are interpreted as descriptive industrial observations rather than as evidence from a randomized or controlled experiment.
The quantitative observations highlight differences in defect rates across development approaches and complexity levels. In particular, ADD consistently showed lower average defects per function, most notably for high-complexity cases. However, these numerical trends do not capture the full picture of the methodology’s adoption and practical use. The following section discusses qualitative insights and contextual factors observed during the 18-month period, complementing the descriptive findings with practical experience from Team~2.

\subsubsection{Observations and Insights}

The quantitative analysis presented in Table~\ref{tab:team2} and Table~\ref{tab:bugs_comparison} shows several patterns regarding the relationship between development methodology, function complexity, and defect occurrence.

First, ADD was associated with lower defect rates across all complexity levels. The reduction was most evident for high-complexity functions, where the average was 1 bug per function compared to 4 for TDD and 9 for the test-last approach. This suggests that the structured modeling and systematic test derivation used in ADD may help mitigate risks introduced by complex control flows and multiple decision points.

Second, the data suggest that ADD adoption was associated with a smaller observed gap between experience levels. To support this observation, Table~\ref{tab:bugs_per_function_experience} reports the average number of bugs per delivered function by experience level, development approach, and complexity level. This breakdown is important because function complexity was not evenly distributed across experience groups, making aggregate comparisons between junior and senior developers potentially misleading.

\begin{table}[ht]
\centering
\caption{Average bugs per delivered function by experience level, method, and complexity (Team~2)}
\label{tab:bugs_per_function_experience}
\renewcommand{\arraystretch}{1.3}
\begin{tabular}{l rrr rrr rrr}
\toprule
\multirow{2}{*}{\textbf{Experience}} 
  & \multicolumn{3}{c}{\textbf{Test-last}} 
  & \multicolumn{3}{c}{\textbf{TDD}} 
  & \multicolumn{3}{c}{\textbf{ADD}} \\
\cmidrule(lr){2-4} \cmidrule(lr){5-7} \cmidrule(lr){8-10}
  & \textbf{Low} & \textbf{Med} & \textbf{High} 
  & \textbf{Low} & \textbf{Med} & \textbf{High} 
  & \textbf{Low} & \textbf{Med} & \textbf{High} \\
\midrule
Junior (1--3 yrs)    & 2.80 & 4.00 & 10.00 & 1.20 & 2.00 & 6.00 & 0.00 & 0.40 & 1.43 \\
Mid-level (4--7 yrs) & 3.00 & 4.00 & 10.00 & 0.86 & 1.43 & 4.67 & 0.00 & 0.55 & 1.00 \\
Senior (8+ yrs)      & 3.33 & 3.00 &  8.00 & 0.67 & 1.00 & 3.58 & ---  & 0.33 & 0.60 \\
\bottomrule
\end{tabular}

\vspace{0.5em}
\begin{minipage}{0.95\textwidth}
\footnotesize
\textit{Note.} Values represent average bugs per delivered function. 
--- indicates that no functions were delivered for this group.
\end{minipage}

\end{table}

Table~\ref{tab:bugs_per_function_experience} shows that lower bug rates under ADD were observed across experience groups and complexity levels. This pattern is consistent with qualitative observations from Team~2, where flowcharts and derived acceptance tests appeared to help less experienced developers reason more systematically about execution paths and edge cases. Informal feedback indicated that junior developers may also have integrated ADD more readily because they had fewer preconceived notions about the development process and less initial skepticism regarding the methodology's value. However, this observation should be interpreted descriptively, as the data were collected in an operational setting and were not produced by a controlled comparison of developers by experience level.

Beyond defect counts, a qualitative review of bug reports revealed additional insights. In the test-last approach, most defects were technical, including logic errors, incorrect branching, or missed edge cases. With TDD, bugs more often stemmed from misunderstood or poorly defined requirements, resulting in functionally incorrect behavior despite technically sound implementations. In contrast, defects found with ADD were less frequent and more diverse. They occasionally related to technical issues, such as logic gaps, or functional ones, such as incomplete or ambiguous flowchart descriptions. These observations suggest that algorithm design quality plays a central role in preventing both categories of defects. While this analysis was not supported by formal classification or statistical validation, the trend was noted consistently across multiple developers and features.

Third, within Team~2, sustained and correct application of ADD appeared to require regular reinforcement. When regular algorithm review meetings and peer feedback sessions were maintained, developers were more likely to apply the methodology consistently. In periods where such reinforcement was reduced, partial or inconsistent use was observed, with a corresponding decrease in the potential benefits.

In summary, these observations point to both technical and organizational implications of ADD adoption in this context. Technically, the methodology contributed to robust implementations, particularly in scenarios involving complex logic. Organizationally, it supported the development of junior team members, provided that appropriate training, coaching, and monitoring mechanisms were maintained.

\subsection{Applicability and Limitations}
\label{sec:limitations}

The results suggest that ADD is particularly applicable in development contexts where expected behavior can be explicitly modeled as execution paths. This includes new feature development involving complex business logic, API orchestration, and migration workflows where correctness, traceability, and regression prevention are important. The findings also suggest that ADD may be introduced incrementally in existing systems, as illustrated by Team 2, which progressively adopted the method during the redesign of a core service. In such contexts, teams can apply ADD around the components being modified, migrated, or refactored, rather than applying it to an entire codebase at once. By modeling expected behavior before implementation changes, teams can derive acceptance tests that help preserve expected behavior while supporting system evolution.

However, several factors limit the applicability, validity, and generalizability of the findings. The effectiveness of ADD depends on the completeness and accuracy of the algorithmic diagrams; unrepresented edge cases or failure scenarios will neither be designed for nor tested. The benefits of ADD may also be lower for very small changes, simple CRUD operations, exploratory prototypes, or projects with highly unstable requirements, where the upfront modeling effort may outweigh the expected benefits. In such cases, lighter approaches such as conventional TDD, BDD-style collaboration, or exploratory prototyping may be more appropriate.

Adoption and consistent use of ADD also depend on team discipline, testing culture, and willingness to maintain consistency between flowcharts, tests, and code. This can be challenging in large, multi-team projects unless review practices, shared ownership, and coaching are established. The experiments were conducted within a single organizational context and on two development teams of different sizes, potentially limiting the extent to which outcomes can be generalized to other industries, organizational cultures, or codebases of different sizes. Finally, environmental factors, such as configuration differences between test and production systems, can still produce defects that escape detection during controlled testing.

\section{Conclusion and Perspectives}
\label{sec:conclusion}

This study examined key challenges in software development, including complexity management, defect reduction, and delivery predictability. Within the studied industrial context, ADD was applied as a structured methodology that shifts the focus from reactive debugging to proactive algorithmic clarification. By translating client needs into structured flowcharts prior to implementation, ADD supports systematic test derivation and explicit representation of execution logic. 

With respect to the research questions introduced in Section~1, the empirical observations reported in Section~4 provide the following insights. Regarding \textbf{RQ1} (comprehensive coverage of client needs and anticipation of edge cases), the case studies show that algorithmic modeling makes execution paths explicit before coding begins. In the reported examples, additional validation branches and edge cases were identified during the flowchart design phase and incorporated prior to implementation. The systematic derivation of acceptance tests from execution paths establishes traceability between modeled behavior and validation artifacts, supporting comprehensive behavioral coverage within the studied context. Concerning \textbf{RQ2} (relationship between the development approach, software defect rates, and overall product reliability), the quantitative indicators collected across both teams, including defect density measurements and bugs-per-function comparisons across development practices and complexity levels, show lower observed defect levels under ADD within the same organizational environment. These results are based on industrial observations rather than controlled experimentation and are interpreted within the stated limitations. With respect to \textbf{RQ3} (management of delivery timelines and predictability), longitudinal delivery data reported for Team~1 indicate a stable delivery cadence over the observed period, with deviations explicitly attributed to external factors. The limited upfront modeling effort did not negatively impact sprint progression and was associated with reduced late-stage rework, contributing to improved predictability in iterative cycles.

Beyond these research questions, the application of ADD highlights several practical insights. Initiating development with algorithm diagrams enhances clarity in complex systems and encourages rigor before coding begins. While adoption may require adaptation for developers accustomed to traditional workflows, structured guidance and collaborative modeling sessions supported long-term integration in the reported cases. The observed effects appear particularly relevant in systems with intricate logic where maintainability and scalability are critical.

\subsection*{Lessons Learned}

While the preceding sections discussed observations in their respective empirical contexts, the main practical lessons from the industrial application of ADD can be summarized as follows:

\begin{itemize}
    \item Flowcharts should be treated as active design artifacts rather than optional documentation. Their value depends on their completeness and on the discipline with which they are kept aligned with tests and code.

    \item Visual conventions, such as color coding for local logic, external calls, and error branches, helped connect design decisions with testing decisions and supported communication during reviews.

    \item Systematic test derivation must be balanced with test-suite scalability. In practice, the use of equivalence blocks and representative paths helped avoid redundant acceptance tests while preserving coverage of relevant behaviors.

    \item ADD provides the most value in high-complexity or high-risk scenarios, where execution paths, edge cases, API orchestration, and error handling must be clarified before implementation.

    \item ADD appeared to support knowledge transfer, developer rotation, and onboarding by making implementation logic easier to understand without first navigating large portions of production code. It also helped less experienced developers reason more systematically about execution paths and edge cases, although this observation remains descriptive and context-dependent.

    \item Sustained adoption requires active reinforcement. Review meetings, coaching, monitoring of method application, and shared ownership of flowcharts and tests helped ADD become embedded in daily development work.
\end{itemize}

For practitioners, the key takeaway is that ADD should be introduced incrementally, starting with complex or risk-prone components where the benefit of explicit behavioral modeling is likely to outweigh the upfront modeling effort. ADD should therefore be considered not only as a notation or documentation practice, but as a team-level engineering routine requiring continued support until it becomes part of regular development work.

The relevance of ADD may also extend to AI-assisted development workflows. In such workflows, a key challenge is that code or test generation from natural-language prompts may suffer from ambiguity, incomplete behavioral assumptions, or hallucinated implementation details. ADD can help mitigate this risk by shifting the developer's effort from directly writing code or tests to specifying behavior more precisely through flowcharts. Because ADD flowcharts explicitly describe execution paths, decision outcomes, edge cases, and expected error scenarios, they can provide more structured input for AI-assisted test-case generation and, potentially, code generation.

From this perspective, ADD may become increasingly relevant as a specification and verification framework for AI-assisted development. If the algorithm is sufficiently precise, AI tools can use it to generate candidate acceptance tests for each selected execution path, while developers remain responsible for validating these tests and implementing or reviewing the production code. The derived tests then act as a control mechanism for generated or manually written code: the flowchart constrains the expected behavior, and the tests verify that the implementation conforms to it. This direction was not evaluated in the present study, but represents a promising area for future work.

Future work may extend the evaluation of ADD across different domains, organizational settings, and team compositions. The development of dedicated tooling for automated test derivation from algorithm diagrams represents a promising direction, potentially reinforcing consistency and reducing manual effort. Broader assessment across additional quality metrics, including long-term maintainability and evolution costs, would further clarify the applicability of ADD beyond the studied industrial context.



\pagebreak

\bibliographystyle{cas-model2-names}

\bibliography{cas-refs}

\section*{Author Biography}

\hypertarget{philippe_bio}{\textbf{Philippe Jawish}} is a mechatronics engineer and Senior Manager at Dassault Systèmes, specializing in cloud application development for the "\textbf{3D}Experience\textregistered\ platform". Leading a dedicated team, he oversees the design and implementation of software solutions aimed at optimizing collaborative and industrial processes in a cloud environment. His expertise includes cloud application architecture, software life cycle management, and the deployment of innovative technologies to enhance workflows and efficiency within digital collaboration platforms.

\hypertarget{pierre_bio}{\textbf{Pierre Evrard}} is a software development engineer and the technical lead of his team at Dassault Systèmes. He is responsible for implementing key functionalities on the "\textbf{3D}Experience\textregistered\ platform", where he contributes to service definition and specializes in back-end development. As a tech lead, he guides the team in designing and optimizing server-side architectures, ensuring efficient and scalable solutions that align with the platform’s strategic objectives.

\hypertarget{alexandre_bio}{\textbf{Alexandre Lemerle}} is a software development engineer responsible for implementing functionalities within a cloud service on the "\textbf{3D}Experience\textregistered\ platform". Specializing in back-end development, he contributes to service definition and works on designing efficient server-side solutions that support the platform's cloud capabilities and enhance its service offerings.

\hypertarget{adrian_bio}{\textbf{Adrian Genin}} is a software development engineer responsible for implementing functionalities within a cloud service on the "\textbf{3D}Experience\textregistered\ platform". Specializing in front-end development, he contributes to service definition and ensures a high-quality user experience, focusing on intuitive and responsive design to enhance usability within the platform.

\hypertarget{layal_bio}{\textbf{Layal Dergham}} is a software development engineer and Director at Dassault Systèmes. She oversees multiple teams responsible for some of the most critical services on the "\textbf{3D}Experience\textregistered\ platform". Her role involves ensuring robust cloud application architecture, high availability, and scalability to support the platform’s demands. Her expertise in cloud infrastructure and strategic planning plays a key role in maintaining the reliability and efficiency of the platform’s core services.

\hypertarget{severin_bio}{\textbf{Séverin Lanfranchi}} is the Chief Technology Officer (CTO) of Dassault Systèmes "\textbf{3D}Experience\textregistered\ platform". An experienced engineer, he is responsible for the architecture of all Cloud Services of the "\textbf{3D}Experience\textregistered\ platform". He plays an active role in shaping the company’s strategic technology directions. His leadership and technical vision are instrumental in driving innovation and ensuring the platform’s alignment with Dassault Systèmes’ long-term goals.

\end{document}